\documentclass[preprint,aps,amsmath,amssymb,nofootinbib]{revtex4}
\usepackage[utf8]{inputenc}
\usepackage[english]{babel}
\usepackage{amsmath}
\usepackage{amsfonts}
\usepackage{amssymb}
\usepackage{graphicx}
\usepackage{color}
\usepackage{subfigure}

\newcommand{\modif}{}

\begin{document}

\title{Kinetic simulations of the Chodura and Debye sheaths for magnetic fields with grazing incidence}

\author{David Coulette}
\email{coulette@unistra.fr}
\author{Giovanni Manfredi}
\email{manfredi@unistra.fr}
\affiliation{Institut de Physique et Chimie des Mat\'eriaux de
Strasbourg, CNRS and Universit\'e de Strasbourg, BP 43, F-67034 Strasbourg Cedex 2, France }

\date{\today}

\begin{abstract}
When an unmagnetized plasma comes in contact with a material surface, the difference in  mobility between the electrons and the ions creates a non-neutral layer known as the Debye sheath (DS). However, in magnetic fusion devices, the open magnetic field lines intersect the structural elements of the device with near grazing incidence angles. The magnetic field tends to align the particle flow along its own field lines, thus counteracting the mechanism that leads to the formation of the DS.
Recent work using a fluid model [P. Stangeby, Nucl. Fusion {\bf 52}, 083012 (2012)] showed that the DS disappears when the incidence angle is smaller than a critical value (around $5^\circ$ for ITER-like parameters). Here, we study this transition by means of numerical simulations of a kinetic model both in the collisionless and weakly collisional regimes. We show that the main features observed in the fluid model are preserved: for grazing incidence, the space charge density near the wall is reduced or suppressed, the ion flow velocity is subsonic, and the electric field and plasma density profiles are spread out over several ion Larmor radii instead of a few Debye lengths as in the unmagnetized case. As there is no singularity at the DS entrance in the kinetic model, this phenomenon depends smoothly on the magnetic field incidence angle and no particular critical angle arises. The simulation results and the predictions of the fluid model are in good agreement, although some discrepancies subsist, mainly due to the assumptions of isothermal closure and diagonality of the pressure tensor in the fluid model.
\end{abstract}

\maketitle


\section{Introduction}
In magnetic fusion devices such as tokamaks, the confining magnetic field is designed so that the field lines
that intersect some machines components do so with near grazing incidence in order to maintain power deposition within
sustainable limits. Due to the large difference in inertia between the ions and the electrons, the latter tend to be lost to the
absorbing wall faster than the former, leading to the formation of a thin (a few Debye lengths wide) positively-charged
transition layer in front of the wall, the so-called Debye sheath (DS) (see \cite{Robertson2013} for a large-scope review on the topic). The
resulting large electric field in the DS repels the electrons and accelerates the ions, leading to a sustainable steady-state with zero
net current at the wall.

In the presence of a magnetic field whose direction is not normal to the wall,
the structure of the transition is more intricate. The magnetic field maintains the ions flow aligned with its own direction, while the electric field tends to accelerate them normally to the wall, leading to a competition between these two effects.
In the case of nearly grazing incidence, the particle motion
along the normal to the wall is essentially cyclotronic, resulting in a strongly reduced net flow in that direction. The
efficiency of the confinement decreases when one approaches the wall, as more and more Larmor orbits intersect the
wall. As the electrons are more strongly confined than the ions, there exists a new transition layer, the so-called Chodura sheath (CS) or magnetic pre-sheath \cite{Chodura82}, where the imbalance between the ionic and electronic flows is sufficiently
compensated by the difference in confinement to maintain quasi-neutrality. This transition layer, between a fully
magnetized plasma flow and the wall is typically a few ion Larmor radii thick. Since generally $\rho_i \gg \lambda_D$ the plasma-wall transition is globally smoother
than in the purely electrostatic case, with smaller spatial gradients for the electric field and plasma density near the wall.

In the most general case, the DS and the CS coexist: the imbalance
between the ionic and electronic parallel flow still requires the formation of a positively charged DS in order to ensure ambipolarity at the wall.
The boundary between the CS and the DS is characterized by the breakdown of quasi-neutrality and the onset of a supersonic ion flow velocity at the entrance of the DS. For unmagnetized plasmas, this reduces to the well-known Bohm criterion \cite{bohm1949,Riemann91}.
A similar criterion was derived by Chodura \cite{Chodura82} in the magnetized case, which requires the \emph{parallel} ion flow velocity at the entrance of the CS to be supersonic.

In the landmark study by Chodura \cite{Chodura82}, the main features of the CS-DS transition were described using both a fluid model and numerical results from particle-in-cell (PIC) simulations. Further studies of the plasma-wall transition, focussing on its stability, were performed with PIC simulations \cite{Daube98,Daube99}. The fluid model was later extended with friction terms to encompass both the magnetic and collisional presheath \cite{Riemann94} {\modif (and more recently \cite{Gao2003})}. This model was recently used to show some partial agreement with experimental data \cite{siddiqui2014} in a different regime ($\lambda_{coll} \approx \rho_i \gg \lambda_D$) with respect to the one considered here ($\lambda_{coll} \gg \rho_i \gg \lambda_D$).

In a recent work \cite{Stangeby2012}, Stangeby also used a fluid model to examine the CS-DS transition for low values (a few degrees) of the incidence angle of the magnetic field, ie, in the range relevant to the plasma-divertor interaction in fusion devices. Importantly, this study showed the existence of a critical incidence angle under which the plasma-wall transition occurs without the need for the formation of the DS.
As a result, the electric field and the plasma density gradients are not restricted to the (very thin) DS, but extend much further (a few ion Larmor radii) into the CS. This effect is significant enough to have a non-negligible impact on prompt redeposition of sputtered neutrals in a tokamak scrape-off-layer (SOL) {\modif \cite{Stangeby2000,Chankin}.}

This potentially important application warrants a more detailed analysis of this phenomenon, going beyond the simple fluid approach that was used in Ref. \cite{Stangeby2012}. The main objective of the present paper is to examine the robustness of Stangeby's results by means of numerical simulations of a kinetic model \cite{coulette2014}.
Various effects that can have an impact on the transition will be analyzed in details, such as the magnitude and incidence of the magnetic field, the effect of collisions, and isotopic effects.
Generally speaking, Stangeby's results are confirmed: the DS disappears for small angles of incidence ($1^\circ-5^\circ$), although the transition is not as clear-cut as in the fluid model.
{\modif
Note that we will not consider here the extreme case $\alpha < m_e/m_i \approx 1^\circ$ (for deuterium), for which the ions reach the wall faster than the electrons, and consequently the sheath structure changes considerably \cite{Tskhakaya2003}}.

The present paper is organized as follows: In Sect. \ref{sec:stangeby}, we summarize the results obtained by Stangeby using a fluid model. In Sect. \ref{sec:model},
we describe the kinetic model and the numerical method and parameters.
In Sect. \ref{sec:nocoll}, we examine the CS-DS transition using a collisionless model, with parameters and boundary conditions chosen to match as closely as possible those of Ref. \cite{Stangeby2012}.
In Sect. \ref{sec:fluid_kin_comp}, we directly compare the spatial profiles obtained from the fluid model and kinetic simulations.
In Sect. \ref{sec:coll}, we introduce a collision operator in our kinetic model, and use it to check the robustness of the observations made in the collisionless regime. In section \ref{sec:conclusions}, we summarize the main conclusions of this study and mention some of the key issues that remain to be addressed.

\section{Stangeby's result from fluid theory}
\label{sec:stangeby}

\begin{figure}
\begin{center}
\includegraphics[width=0.48\textwidth]{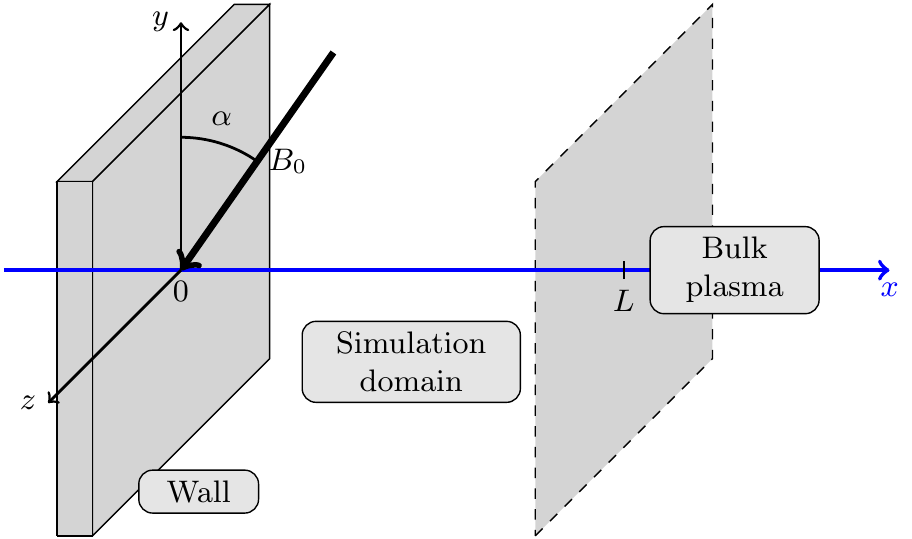}
\caption{Geometry of the model}
\label{fig:geometry}
\end{center}
\end{figure}

Stangeby \cite{Stangeby2012} considered a plasma composed of electrons of charge $-e$ and a single ion species of charge $q_i=Z_i e$. The plasma is bounded by a fully
absorbing wall on one side, while thermal equilibrium is assumed far from the wall (see Fig. \ref{fig:geometry}). Noting $x$ the direction
corresponding to the normal to the wall, the system is assumed invariant by translation in the $(y, z)$ plane parallel to
the wall. The plasma is magnetized by an external magnetic field $B_0$, constant in space and time, whose direction
is normal to $\mathbf{e}_z$ and makes an angle $\alpha$ with the wall, i.e $\mathbf{B}_0 = B_0 (\sin \alpha \mathbf{e}_x + \cos \alpha \mathbf{e}_y)$. The self-consistent magnetic field generated by plasma currents is neglected.

The main result of Ref. \cite{Stangeby2012} is the existence of a critical angle $\alpha_c$ for which there is strictly no Debye sheath, or more precisely the
average flow along the normal to the wall never becomes sonic. We will first reestablish this result with slightly more relaxed assumptions in order to treat both sonic and supersonic regimes, and then examine the actual simulation results. Though the model used in \cite{Stangeby2012} is a fluid one, the result is actually quite generic. From the ion flux conservation $\partial_x j_{xi}=0$ we have :
\begin{equation}
\langle v_x \rangle_i (x) = \frac{j_{xi}^W}{n_i(x)}= \frac{Z_i j_{xi}^W}{n_e(x)+\rho/e},
\end{equation}
{\modif
where $\langle v_x \rangle_i$ is the mean ion velocity, $j_{xi}$ is the mean ion current, $n_{i,e}$ is the ion (electron) number density, and $\rho=e(Z_i n_i- n_e)$ is the charge density.}
The superscript \textit{``W"} refers to the wall and $\langle \cdot \rangle$ stands for the averaging operator over velocity space.
Using the ambipolarity condition at the wall $Z_i j_{xi}^W = j_{xe}^W$ we have
\begin{equation}
\langle v_x \rangle_i (x) = \left[
\langle v_\parallel \rangle_e^W \sin \alpha + \langle \mathbf{v}_\perp \cdot \mathbf{e}_x \rangle_e^W \cos \alpha
\right]
\frac{n_e^W}{n_e(x)+\rho(x)/e}.
\label{eq:ion_flow_eq}
\end{equation}

Now we make two assumptions. The first is on the ratio $\frac{n_e^W}{n_e(x)+\rho(x)/e}$, which we will take to be less than unity. Such condition is fulfilled in the case of a quasi-neutral region ($\rho \approx 0$, as in the CS) or a positively charged region ($\rho > 0$, as in the standard DS), subject to the condition of a decrease of the electron density when
one approaches the wall ($\partial_x n_e \le 0$).  This  is clearly the case for Boltzmann electrons and a negatively charged wall, as was assumed in Ref. \cite{Stangeby2012}. Whatever the exact assumptions, as long as  $\frac{n_e^W}{n_e(x)+\rho(x)/e} \le 1$ we obtain a bound on the ion flow velocity
\begin{equation}
\vert \langle v_x \rangle_i (x) \vert \le \vert
\langle v_\parallel \rangle_e^W \sin \alpha + \langle \mathbf{v}_\perp \cdot \mathbf{e}_x \rangle_e^W \cos \alpha
 \vert.
\label{eq:cs_bound_ion_general}
\end{equation}

The second assumption is that the electrons are perfectly magnetized up to the wall, ie, $\langle \mathbf{v}_\perp \cdot \mathbf{e}_x \rangle_e^W=0$.
This becomes obviously false for distances smaller than the electron Larmor radius
$\rho_e$ from the wall, but can be considered a reasonable approximation as long as the electron flow variation is mild. We then have
\begin{equation}
\vert \langle v_x \rangle_i (x) \vert \le
\sin \alpha \vert \langle v_\parallel \rangle_e^W  \vert.
\label{eq:cs_bound_ion}
\end{equation}
For sufficiently small $\alpha$, the bound of Eq. \eqref{eq:cs_bound_ion} may prevent the ion mean velocity $\langle v_x \rangle_i$ from becoming supersonic, in which case no DS is required to guarantee ambipolarity. This happens when $\alpha$ is equal or smaller than the critical value $\alpha_c$ defined as
\begin{equation}
\sin \alpha_c = \frac{c_s}{\vert \langle v_\parallel \rangle_e^W \vert}.
\end{equation}
In the case of a half-Maxwellian electron parallel velocity distribution at the wall, one has $\langle v_\parallel \rangle_e^W= \sqrt{T_{e0}/(2\pi m_e)}$ and the result of Ref. \cite{Stangeby2012} is readily obtained.
The underlying physical phenomenon is essentially the limitation of the electron current at the wall by the magnetic
field, which entails a limitation of the ion current. For sufficiently small $\alpha$, an ambipolar flow along $x$ can be
maintained at the wall without requiring strong ion acceleration, so that there is no need for a DS.

A few points of importance should be noted:
\begin{enumerate}
\item While the bound on the CS ion flow velocity in Eq. \eqref{eq:cs_bound_ion_general} is quite generic, the notion of a well-defined critical angle stems from two assumptions: a Bohm criterion on the ion velocity for the existence of the sheath (ie, $\vert \langle v_x \rangle_i \vert \ge c_s$ at the sheath entrance) and perfect magnetization of the electrons. In a kinetic model such as the one considered later on in this paper, the relationship between the mean ion flow velocity and the sheath stability is not as direct as the simple Bohm criterion.
\item A second point is the fact that the bound of Eq. \eqref{eq:cs_bound_ion} and the critical angle do not depend explicitly on the flow at the CS entrance, and are thus valid in the CS in both the sonic and supersonic regimes. This is in contrast with the result presented in Appendix A of Ref. \cite{Stangeby2012} which relies on the erroneous use in a supersonic case of the potential drop in the CS that had been established for a sonic case {\modif (Eq. (33) in \cite{Stangeby2012}, used in conjunction with Eq. (A3) of the same paper).}
\item As was noted in \cite{Stangeby2012}, in a model accounting for the finite electron Larmor radius, the angular dependency of the electron current would be more complex than the simple $\sin \alpha$ behaviour considered here.
\end{enumerate}

\section{Kinetic model and numerical parameters}
\label{sec:model}
In the kinetic model considered here, the dynamics of the ions is described by the evolution of the phase-space distribution function $f_i(t,x,v_x,v_y,v_z)$ obeying the collisional Vlasov equation
\begin{equation}
\partial_t f_i + v_x \partial_x f_i + \left( \frac{q_i}{m_i} \mathbf{E} + \omega_{ci} \mathbf{v} \times \mathbf{e}_z \right) \cdot \mathbf{\nabla}_v f_i = C_i (f_i),
\label{eq:vlasov}
\end{equation}
where $\omega_{ci}=Z_i eB_0/m_i$ is the ion cyclotron frequency.
In all results presented hereafter the collision operator, whenever present, is a Bathnagar-Krook-Gross (BGK) linear relaxation operator \cite{BGK1954}, which drives the distribution function to an isotropic Maxwellian distribution, ie, $C_i (f_i ) = -\nu_i (f - f_i^M )$ where $\nu_i$ is the ion relaxation rate
and $f_i^M = n_{i0} \left(\frac{m_i}{2\pi T_{i0}}\right)^{3/2}\exp\left[-\frac{m_iv^2}{2T_{i0}}\right]$. At the wall $(x=0)$, an absorbing boundary condition is assumed in $x$ for the incoming part of the distribution function, ie, $f_i(t,0,v_x,v_y,v_z) = 0$ for $v_x > 0$. On the plasma side ($x = L$), the incoming particle
distribution is prescribed by $f_i(t,L,v_x,v_y ,v_z) = f_i^{in}(v_x,v_y,v_z )$ for $v_x < 0$. In the collisional simulations $f_i^{in}$ is simply a
Maxwellian with bulk plasma parameters (the same that is used for the BGK operator).
In the collisionless simulations, it is a field-aligned drifting distribution with parallel velocity that satisfies the Chodura criterion at the CS entrance (see Sect. \ref{sec:bc_collisionless}).

The electrostatic field $\mathbf{E} = -\partial_x \phi \,\mathbf{e}_x$ is computed from the electrostatic potential by solving the Poisson equation
\begin{equation}
\partial_{xx}^2  + \frac{e}{\epsilon_0}(Z_i n_i -n_e)=0
\label{eq:Poisson}
\end{equation}
with a Dirichlet boundary condition $\phi=0$ at $x=L$ and a Von Neumann condition $E_x= \sigma/\epsilon_0$ at the wall. The wall charge
surface $\sigma$ is computed by integrating in time the outgoing net electric current:
\[\displaystyle \sigma= - e\int\limits_0^{t} \sum_{s=i,e} Z_s j_{xs}(t',x=0) dt',
\]
with
$j_{xs} = \int v_x f_s d^3v$. The full electron kinetic dynamics is not resolved, but instead a Boltzmann law is assumed for the electron density $n_e = n_{ref} \exp\left[e(\phi-\phi_{ref} )/T_{e0})\right]$. The reference quantities are defined at $x=L$ by $\phi_{ref}=0$ and $n_{ref}= n_i(L)$. The outgoing electron flux at the wall is computed by assuming a half-Maxwellian distribution and is given by
\begin{equation}
j_{xe}^W = - \sin \alpha \sqrt{\frac{T_{e0}}{2\pi m_e}} n_{ref} \exp \left[ e(\phi-\phi_{ref})/T_{e0}\right].
\end{equation}
The latter relation does not take into account finite electron Larmor radius effects, as it is assumed that $j_{xe}= \sin \alpha j_{\parallel e}$.

All numerical simulations were performed using the Eulerian code described in Ref. \cite{coulette2014}. The numerical scheme is based on a split-operator technique for the time-stepping algorithm, with interpolations performed with a positive flux conservative (PFC) scheme \cite{Filbet2001}.
In all cases, starting from a uniform  Maxwellian plasma, the system is left to relax towards a stationary state. A first set of simulations were run in a collisionless regime ($\nu_i=0$) over a spatial domain limited to the CS+DS region, covering a few ion Larmor radii. A second set of simulations were run in a collisional regime where the full transition from an isotropic Maxwellian plasma to the wall is considered, including the collisional presheath.
In both cases, parametric scans with  $\alpha \in \left\lbrace 2^{\circ},3^{\circ},4^{\circ},5^{\circ},10^{\circ},15^{\circ},30^{\circ},45^{\circ},60^{\circ},90^{\circ} \right\rbrace$ were performed.

\section{Collisionless plasma-wall transition}
\label{sec:nocoll}

The parameters of the first set of simulations were set in order to match as closely as possible those of the fluid model
used in Ref. \cite{Stangeby2012}. {\modif  The simulation box length is between $L \approx 120 \lambda_D$ and $L \approx 800 \lambda_D$, depending on the strength of the magnetic field
(in Stangeby's quasi-neutral model, since the Debye length vanishes, the CS entrance is rejected at infinity)}.
Parametric scans in the incidence angle $\alpha$ were performed for hydrogen ($m_i=m_H$) and deuterium ($m_i=2 m_H$). The magnetic field intensity is such that $\omega_{ci}=0.05 \omega_{pi}$, ie  $\rho_i = 20 \lambda_D$.
{\modif
For all simulations, we assumed equal temperatures $T_{i0} = T_{e0}$}.
For brevity, the local value of any quantity $X$ expressed at the wall ($x=0$) and at the magnetic presheath entrance ($x=L$) will be tagged respectively as $X^W$ and $X^{CSE}$.

\subsection{Boundary conditions}
\label{sec:bc_collisionless}
At the plasma boundary, ie, the CS entrance,
the incoming ion flux {\modif should be supersonic (Chodura criterion) and aligned with the magnetic field direction. To this end, we prescribe the following
distribution function at $x = L$:}
\begin{equation}
f_i^{in}= K H(-v_\parallel) \left(\frac{\vert v_\parallel \vert}{v_{thi}}\right)^{\beta} \exp\left[ -\frac{ \vert v\vert^2}{2 v_{thi}^ 2}\right],
\label{eq:fin}
\end{equation}
where $H$ is the Heaviside function, $v_{thi}= \sqrt{T_{i0}/m_i}$, $K= \frac{n_{i0}}{2\pi v_{thi}^3} 2^{\frac{1-\beta}{2}} \Gamma(\frac{\beta +1}{2} )$, with $\Gamma$ the Euler gamma function. The average parallel velocity corresponding to  $f_{i}^{in}$ is $\langle v_\parallel \rangle= -v_{thi} \sqrt{2}\Gamma(\frac{\beta +2}{2} )/\Gamma(\frac{\beta +1}{2})$. In the results presented in this section the $\beta$ exponent was set equal to $2$, leading to an average flow $\langle v_\parallel \rangle= - 1.6 v_{thi}$, {\modif ie, slightly supersonic. Smaller values of the parallel velocity may run the risk of destabilizing the transition.
The above distribution is compatible with fluid models that assume a sonic or slightly supersonic flow at the entrance of the CS. In addition, it is also compatible with the velocity distribution obtained self-consistently from a kinetic model that incorporates weak collisions, as we shall show in Sect. \ref{sec:coll}.
}

\begin{figure*}
\begin{center}
\subfigure[]{\includegraphics[width=0.45\textwidth]{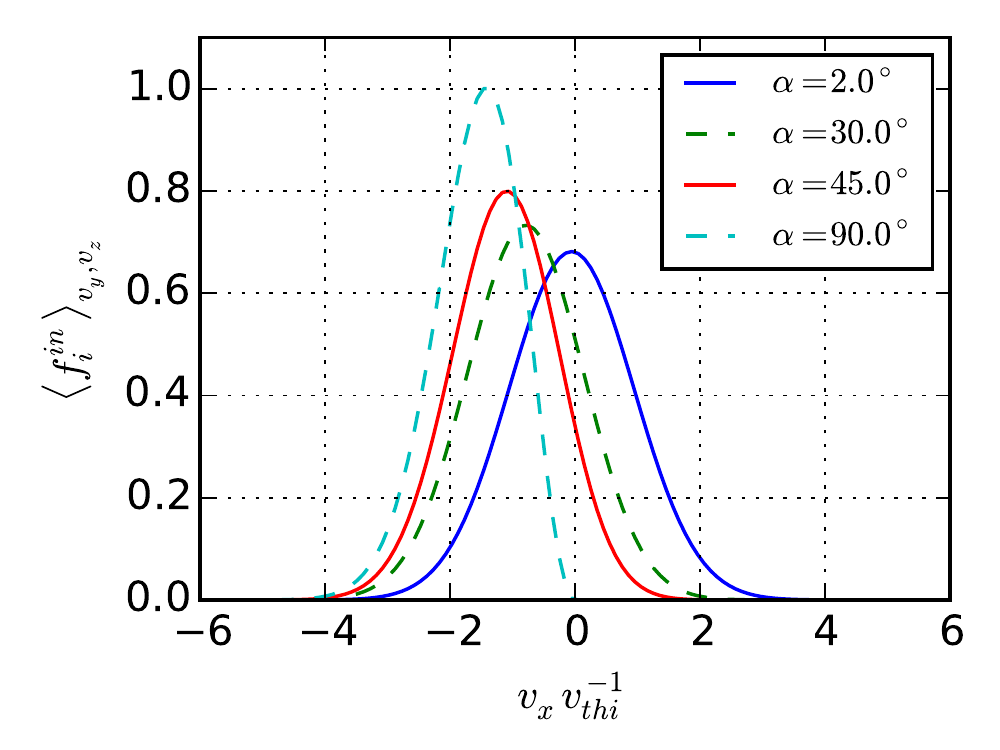}}
\subfigure[]{\includegraphics[width=0.45\textwidth]{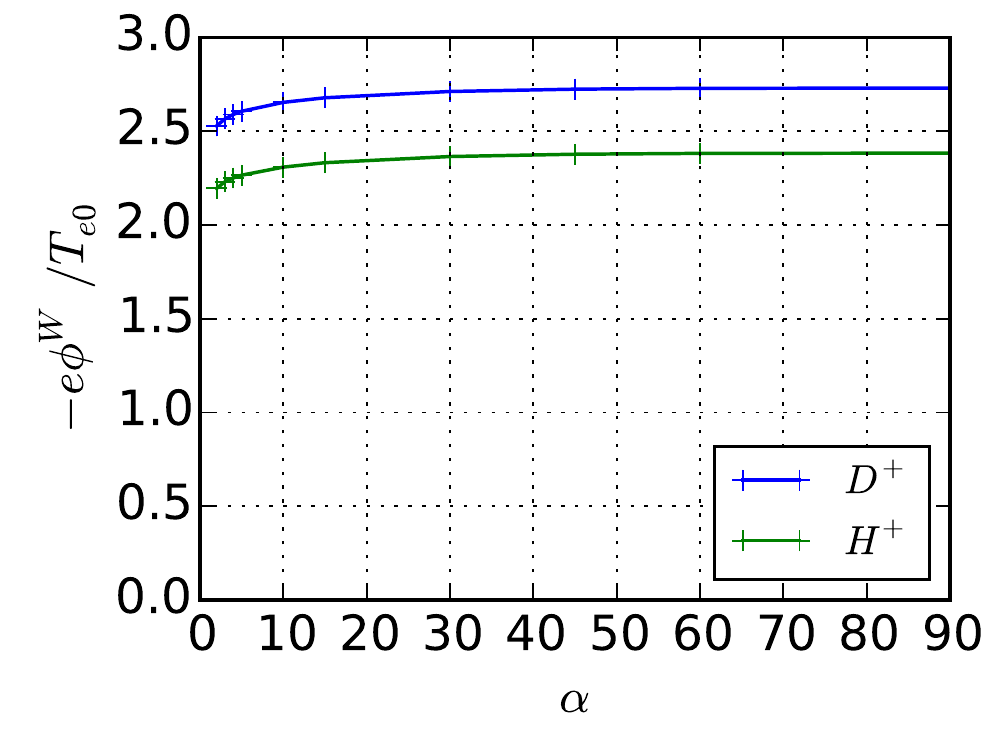}}
\caption{(a) Prescribed ion distribution functions at $x = L$  averaged over $(v_y, v_z)$, for a few values of the incidence angle $\alpha$ (the distributions are normalized to the peak value of the $\alpha=90^\circ$ case); (b) Total potential drop in the CS and DS as a function of $\alpha$.}
\label{fig:f_in_alpha_phi_wall_nocoll}
\end{center}
\end{figure*}

In Fig. \ref{fig:f_in_alpha_phi_wall_nocoll}a the $v_x$ dependency of the incoming distribution function is shown for a few values of $\alpha$. The case $\alpha=90^{\circ}$ corresponds to $v_x=v_\parallel$. One should note that the parallel velocity distribution is not a Maxwellian, and that its effective "temperature" $T_i^{in \parallel} = P_{\parallel\parallel}^{in}/n \approx 0.45 T_{i0}$ is smaller than $T_{i0}$.
In a magnetic-field-aligned basis such as $(\mathbf{b},\mathbf{e}_z \times \mathbf{b},\mathbf{e}_z)$, the kinetic pressure tensor is diagonal but anisotropic. In the $(x,y,z)$ basis, it is not even diagonal anymore and the  various  components of the pressure tensor vary with $\alpha$. For instance, the $xx$ component of the pressure tensor,
for $\beta=2$, is equal to $P_{xx}^{in} \approx n_{i0} T_{i0} (1-0.55 \sin^2 \alpha)$.

{\modif
In a collisionless model, the total potential drop from the CS entrance to the wall is independent on the angle $\alpha$. However, for very small angles, numerical errors (due to the presence of a small but non-zero electric field near $x = L$) slightly break this invariance. This entails a small variation with $\alpha$ of the total potential drop, as seen in Fig. \ref{fig:f_in_alpha_phi_wall_nocoll}b.
However, this small error does not affect the main physical conclusions that can be drawn from the forthcoming numerical simulations.
}

\subsection{Effect of the angle of incidence}

\begin{figure*}
\subfigure[]{\includegraphics[width=0.45\textwidth]{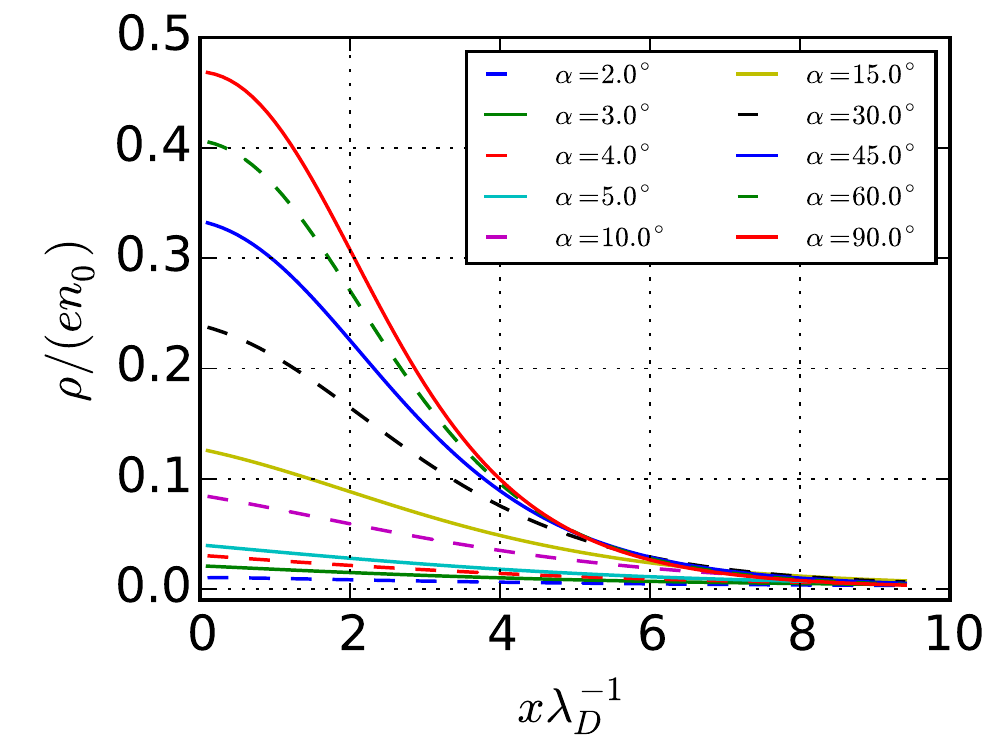}}
\subfigure[]{\includegraphics[width=0.45\textwidth]{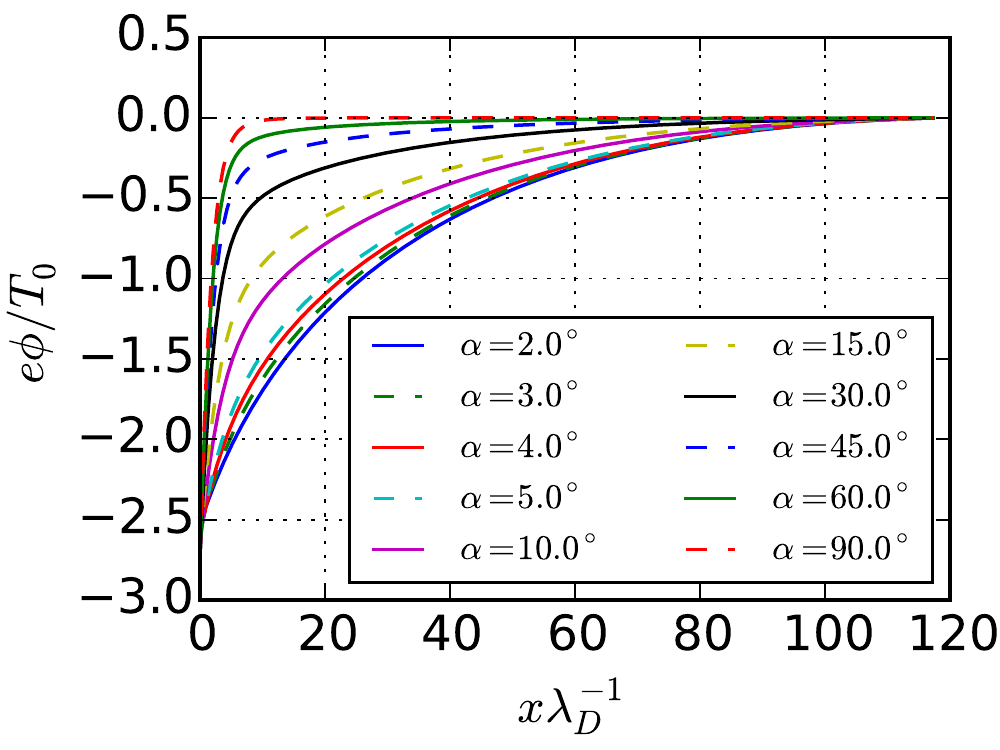}}
\caption{Spatial profiles of the charge density (a) and the electric potential (b), for a collisionless case with deuterium ions.}
\label{fig:rho_phi_profile_nocoll_D}
\end{figure*}

\begin{figure*}
\subfigure[]{\includegraphics[width=0.45\textwidth]{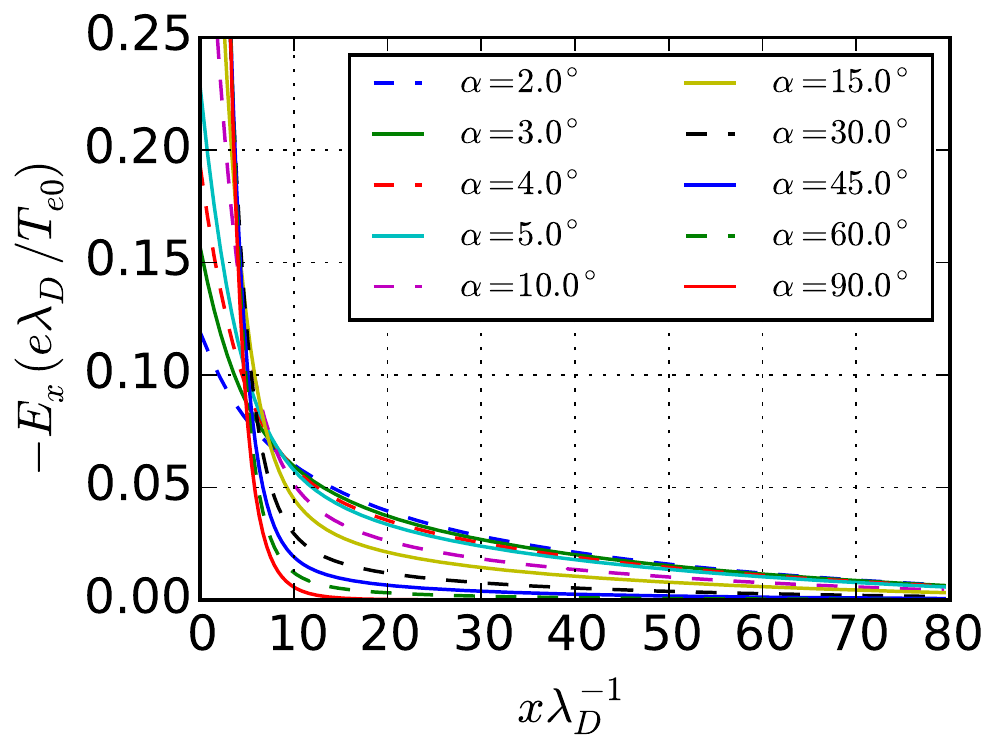}}
\subfigure[]{\includegraphics[width=0.45\textwidth]{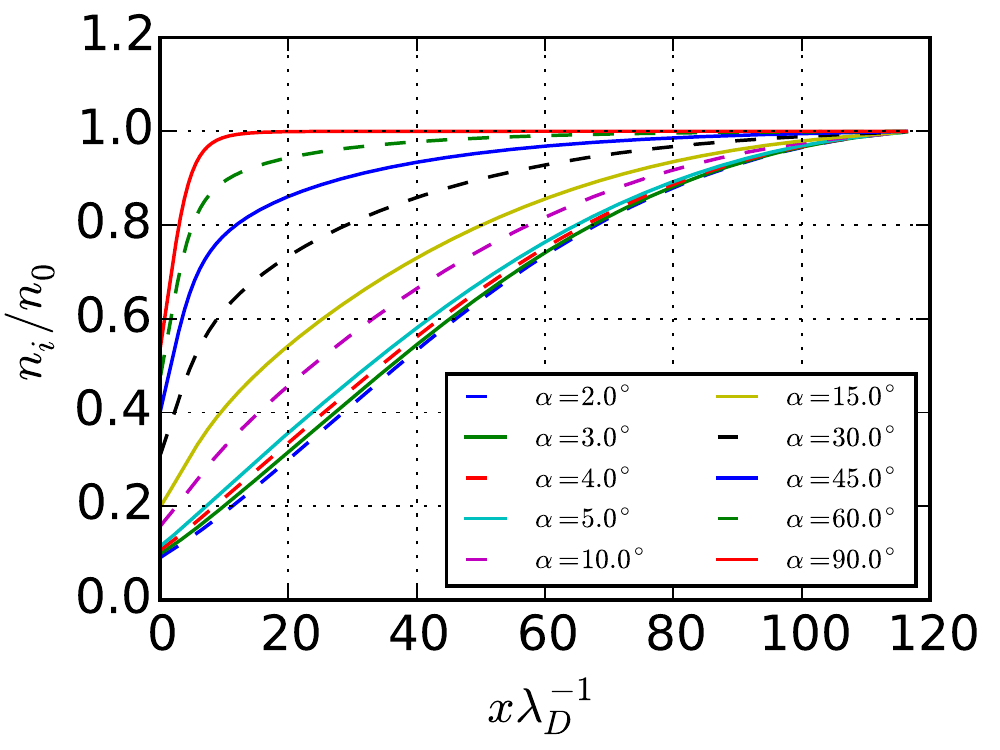}}
\caption{Spatial profiles of the electrostatic field (a) and the ion density (b), for a collisionless case with deuterium ions.}
\label{fig:Ex_dens_ion_profile_nocoll_D}
\end{figure*}

We will now consider the parametric dependency of the CS-DS transition with the magnetic field incidence
angle $\alpha$. Figure \ref{fig:rho_phi_profile_nocoll_D}a shows that the space charge density near the wall decreases rapidly with decreasing $\alpha$.
Although the charge density does not strictly vanish (nor changes sign), the strong limitation of the space charge density is a clear signature that the DS progressively disappears at small incidence angles.
In addition, the spatial profile of the electric potential (Fig. \ref{fig:rho_phi_profile_nocoll_D}b) evolves
from a two-scale profile at large $\alpha$ -- typical of the CS-DS transition -- to a smooth evolution at low $\alpha$.
As a consequence, although the peak of the electric field decreases strongly as the DS vanishes (Fig. \ref{fig:Ex_dens_ion_profile_nocoll_D}a), its extension reaches much further into the plasma, several ion Larmor radii from the wall.
As discussed in Ref. \cite{Stangeby2012}, this is of significant importance for the estimation of the prompt redeposition of sputtered impurity ions: indeed, while the overall electric field intensity decreases with $\alpha$, it will affect
sputtered neutrals ionized farther from the wall and increase prompt redeposition.

The ion (and thus plasma) density drop is also spread out and reaches lower values with decreasing $\alpha$ (Fig. \ref{fig:Ex_dens_ion_profile_nocoll_D}b). This depletion of the plasma density near the wall (for regions such that $x \le \rho_i$) entails a lower ionization rate for sputtered neutrals, thus lowering prompt redeposition.

\begin{figure}
\begin{center}
\includegraphics[width=0.45\textwidth]{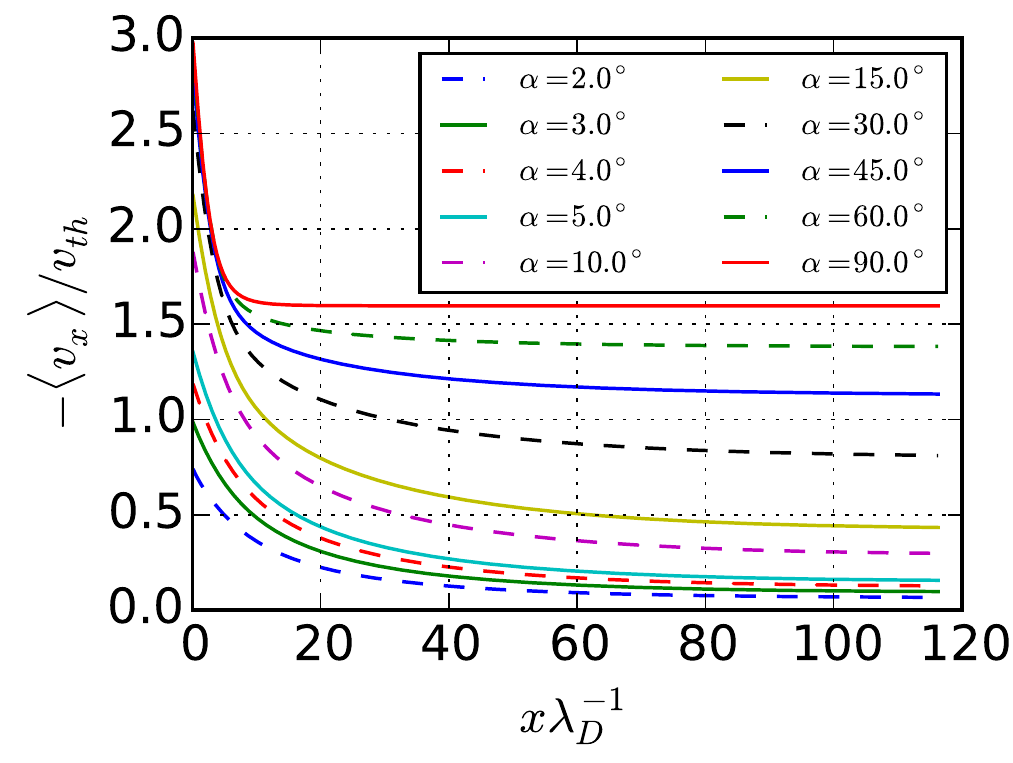}
\caption{\modif Spatial profiles of the average ion flow velocity perpendicular to the wall, for a collisionless case with deuterium ions.}
\label{fig:vx_profile_nocoll_deuterium}
\end{center}
\end{figure}

Let us now consider the ion mean flow perpendicular to the wall (Fig. \ref{fig:vx_profile_nocoll_deuterium}). Due to both the anisotropic nature
of the kinetic pressure tensor and the non-uniformity of the "temperatures" (see Sect.  \ref{sec:fluid_kin_comp} for a discussion of the fluid
closure), we refrain here from normalizing the flow to the usual sound speed $c_s = \sqrt{(T_{i0} + T_{e0})/m_i} \approx 1.4v_{thi}$, which is
strictly valid only in the case of an isothermal closure for the $P_{xx}$ component of the kinetic pressure tensor. In our case, the sound speed can be roughly estimated (from $f_i^{in}$) as ranging from $1.2v_{thi}$ to $1.4 v_{thi}$ when $\alpha$ ranges
from $90^{\circ}$ to $2^{\circ}$, and is very close to $1.4 v_{thi}$ for the lowest range ($\alpha < 15^{\circ}$) of angles considered.

\begin{figure*}
\subfigure[$\alpha=3^{\circ}$]{\includegraphics[width=0.33\textwidth]{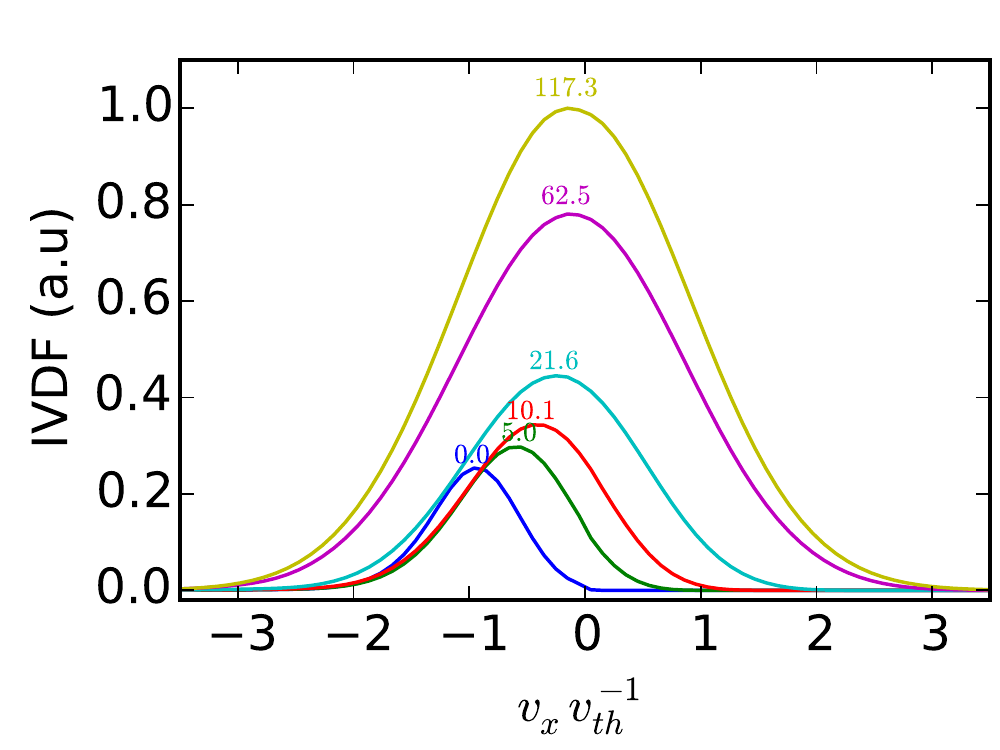}}
\subfigure[$\alpha=30^{\circ}$]{\includegraphics[width=0.33\textwidth]{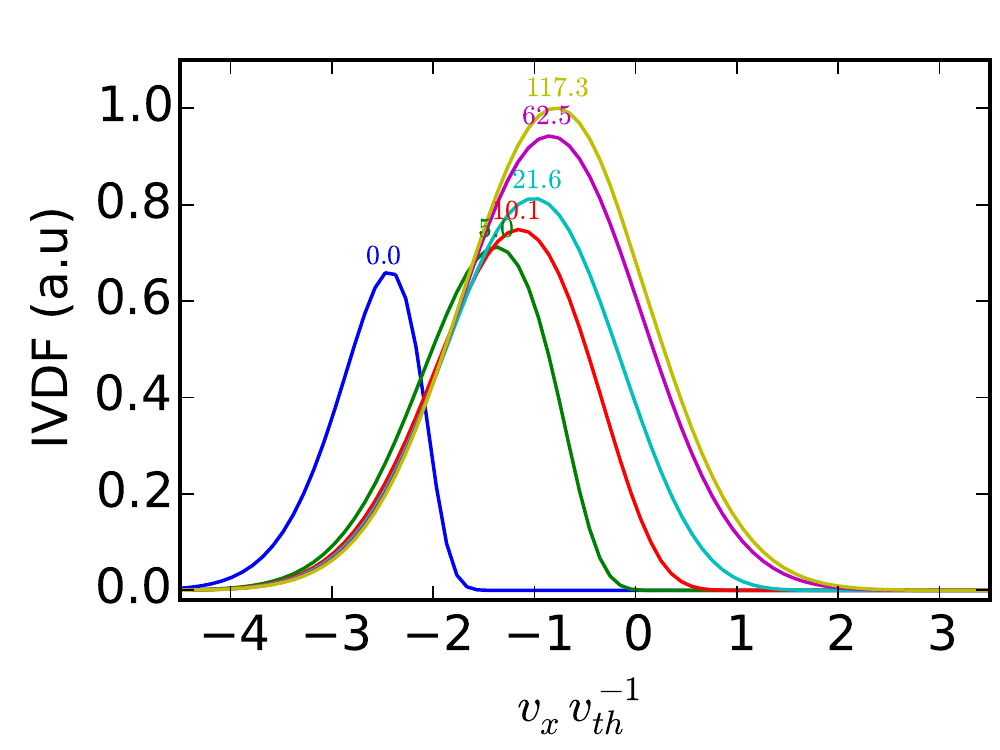}}
\subfigure[$\alpha=90^{\circ}$]{\includegraphics[width=0.33\textwidth]{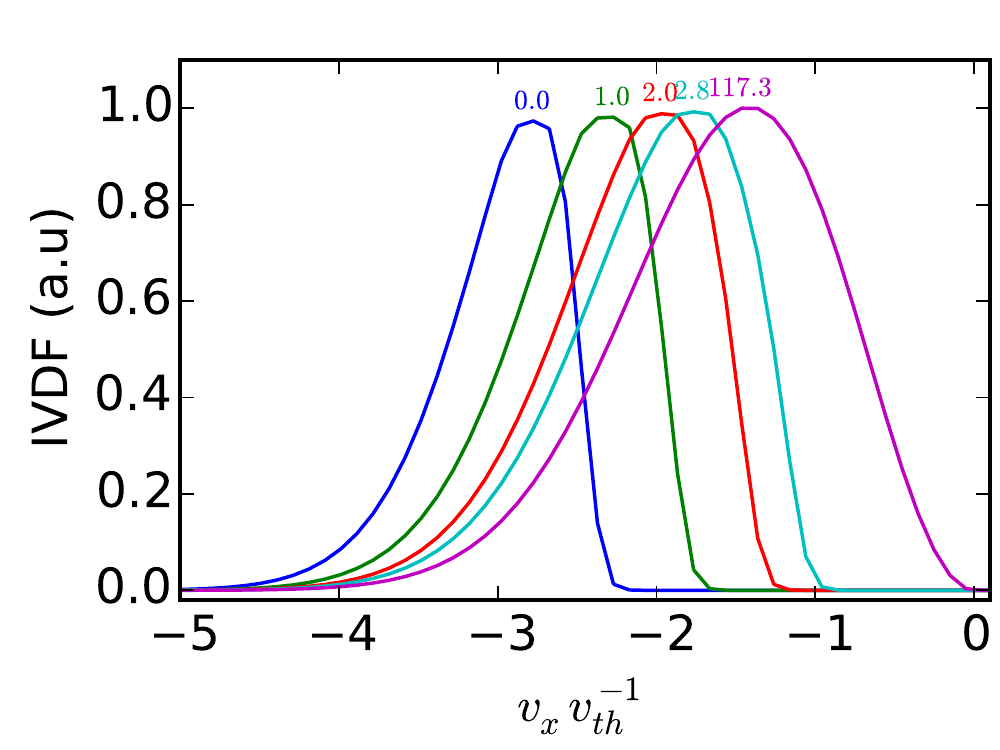}}
\caption{Ion velocity distribution function in $v_x$ for several positions (indicated on top of each curve, in units of $\lambda_D$)
and three values of $\alpha$: (a) $\alpha=3^\circ$, (b) $\alpha=30^\circ$, and (c) $\alpha=90^\circ$. For each value of $\alpha$, all distributions are normalized to their peak value at the entrance of the CS ($x=117.3 \lambda_D$).}
\label{fig:f_vx_nocoll_D}
\end{figure*}

Figure \ref{fig:vx_profile_nocoll_deuterium} clearly shows that {\modif the peak (absolute) value of the ion mean velocity decreases with decreasing $\alpha$ } and is limited to subsonic values for low angles of incidence, below approximately $5^\circ$. Together with the disappearance of the space charge in front of the wall (Fig. \ref{fig:rho_phi_profile_nocoll_D}a), these results confirm Stangeby's conclusion that no DS forms below a certain angle of incidence.
The limitation of both the ion density and the average velocity with decreasing $\alpha$ are clearly visible when examining directly the $v_x$ velocity profile of the ion distribution function (averaged over $v_y$ and $v_z$), as shown in Fig. \ref{fig:f_vx_nocoll_D}.

\begin{figure}
\begin{center}
\includegraphics[width=0.45\textwidth]{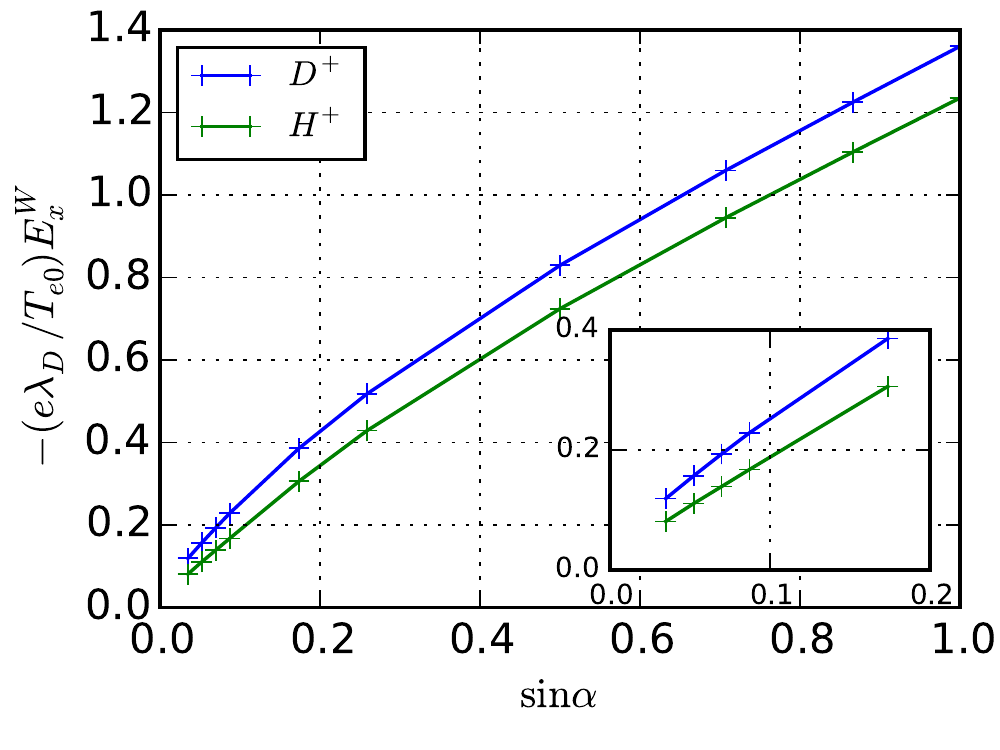}
\caption{Electric field magnitude at the wall as a function of the incidence angle, for a collisionless case with hydrogen or deuterium ions. The inset is a zoom at small angles. }
\label{fig:ex_wall_vs_alpha_nocoll}
\end{center}
\end{figure}

We will now examine more closely the behaviour with $\alpha$ of a few important quantities measured at the wall. The $x$
component of the electrostatic field at the wall is shown in Fig. \ref{fig:ex_wall_vs_alpha_nocoll} as a function of $\sin \alpha$. As expected
from the above observations, it is an increasing function of $\alpha$. For the smallest angles $\alpha\in\lbrace 2^{\circ}, 3^{\circ}, 4^{\circ}, 5^{\circ}, 10^{\circ}\rbrace$ (inset of Fig. \ref{fig:ex_wall_vs_alpha_nocoll}), the evolution is roughly linear in $\sin \alpha$, but the overall behaviour for the full range of angles is less obvious.

\begin{figure*}
\begin{center}
\subfigure[]{\includegraphics[width=0.45\textwidth]{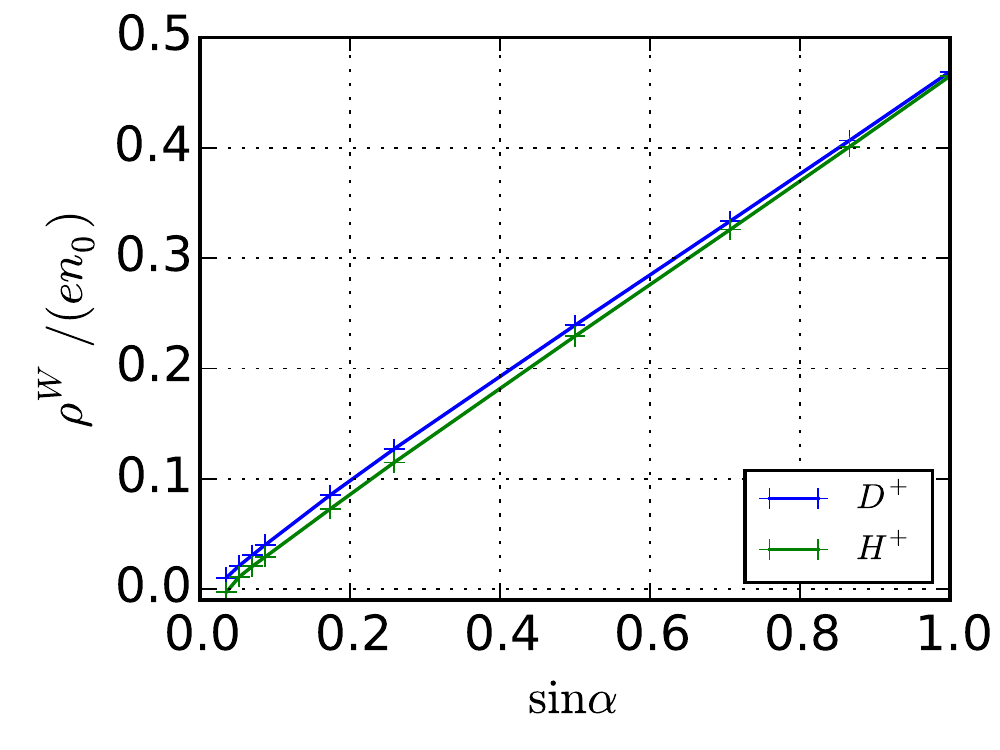}}
\subfigure[]{\includegraphics[width=0.45\textwidth]{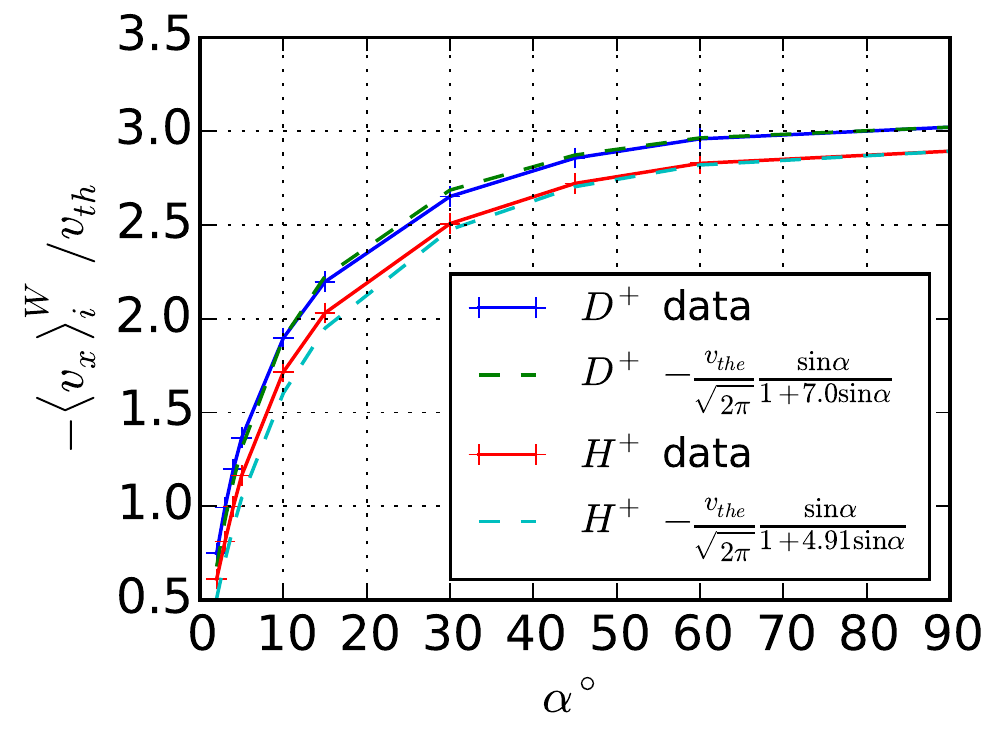}}
\caption{Charge density (a)  and ion mean velocity (b) at the wall as a function of the incidence angle. The coefficient $\kappa$ of Eq. \eqref{eq:vx_wall_fit} is obtained from the simulation data for $\alpha=90^{\circ}$.
}
\label{fig:rho_vx_wall_vs_alpha_nocoll}
\end{center}
\end{figure*}

The space charge density at the wall clearly exhibits a linear dependency
in $\sin \alpha$ (Fig. \ref{fig:rho_vx_wall_vs_alpha_nocoll}a).
This fact allows us to obtain a semi-empirical fit for the ion perpendicular velocity at the wall. Indeed, taking Eq. \eqref{eq:ion_flow_eq} at the wall with an electron current $j_{xe}^W = -\sin \alpha (v_{the}/\sqrt{2\pi}) n_e^W$ we obtain
\begin{equation}
\vert \langle v_x \rangle_i^W \vert = \frac{v_{the}}{\sqrt{2\pi}} \frac{\sin \alpha}{1+\rho^W/(en_e^W)}= \frac{v_{the}}{\sqrt{2\pi}} \frac{\sin \alpha}{1+\kappa \sin \alpha},
\label{eq:vx_wall_fit}
\end{equation}
where $\kappa$ is a fitting parameter. To obtain Eq. \eqref{eq:vx_wall_fit}, we have assumed that $\rho^W \propto \sin \alpha$ (see Fig.  \ref{fig:rho_vx_wall_vs_alpha_nocoll}a) and that $n_e^W$ is independent of $\alpha$.
An interesting fact here is that the coefficient $\kappa$ can be computed
in the normal incidence case ($\alpha=90^{\circ}$), which does not require a full 1D3V model but only a far simpler 1D1V simulation. Once $\kappa$ has been determined, the ion perpendicular flow {\em for any incidence angle} can be computed using Eq. \eqref{eq:vx_wall_fit}.
Some examples of this semi-empirical fit are shown in Fig.  \ref{fig:rho_vx_wall_vs_alpha_nocoll}b, for both hydrogen and deuterium ions.

\subsection{Effect of the magnetic field amplitude at fixed angle ($\alpha = 2^\circ$)}
\begin{figure}
\begin{center}
\includegraphics[width=0.45\textwidth]{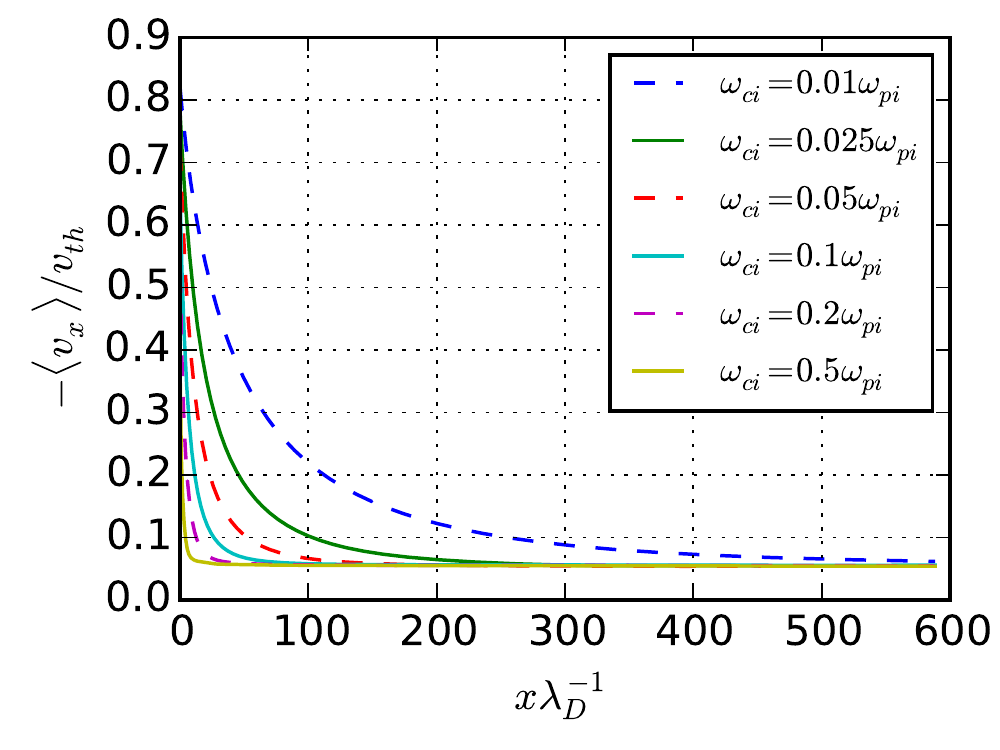}
\caption{Ion flow velocity perpendicular to the wall for various amplitudes of the magnetic field $B_0$ and $\alpha=2^{\circ}$. Collisionless simulations with deuterium ions.}
\label{fig:ion_vx_scanB0}
\end{center}
\end{figure}

\begin{figure}
\begin{center}
\includegraphics[width=0.45\textwidth]{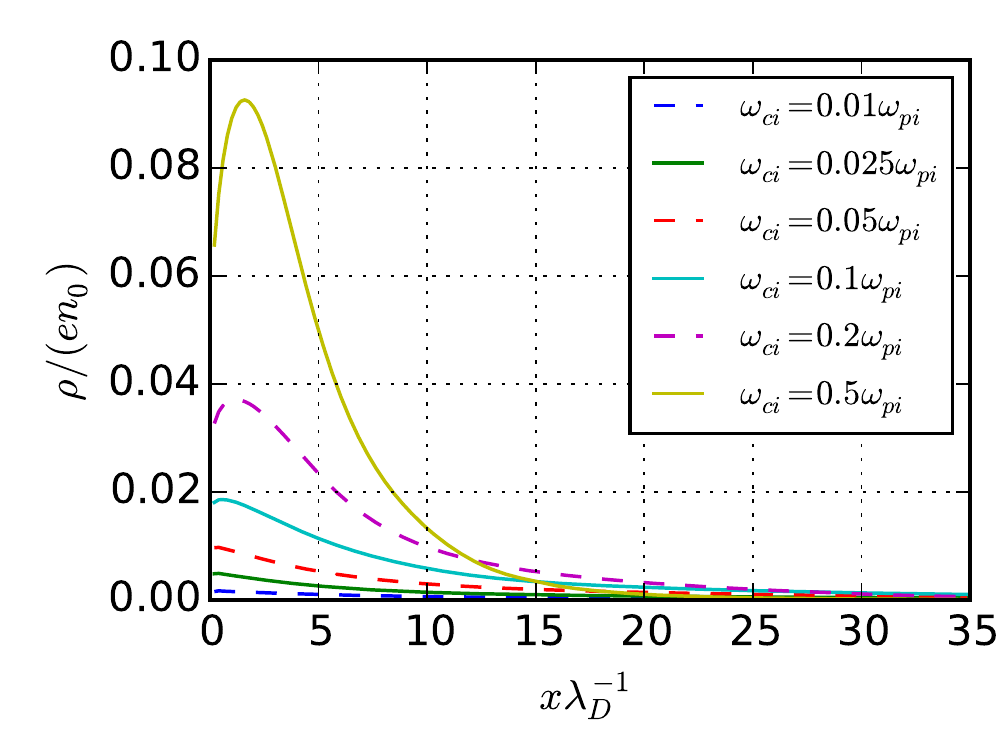}
\caption{Charge density profile for various amplitudes of the magnetic field $B_0$ and $\alpha=2^{\circ}$. Collisionless simulations with deuterium ions.}
\label{fig:rho_scanB0}
\end{center}
\end{figure}

In the simulations considered so far, the scaling $\rho_i=20 \lambda_D \gg \lambda_D$
(or $\omega_{ci}/\omega_{pi} = 0.05 \ll 1$)
was valid. In that regime, decreasing the magnetic field intensity $B_0$ will essentially result in a rescaling of the CS, whose thickness increases with growing $\rho_i$.
{\modif
This is clear from Fig. \ref{fig:ion_vx_scanB0}, for instance in the case $\omega_{ci}=0.01 \omega_{pi}$ ($\rho_i=100 \lambda_D$), where the velocity profile stretches out to several hundred Debye lengths. At the same time, the charge separation near the wall decreases with decreasing magnetic field, and almost disappears for $\omega_{ci}=0.01 \omega_{pi}$ (Fig. \ref{fig:rho_scanB0}). This is because, the CS being larger, the whole potential drop can more easily occur within the CS, with hardly any need for a non-neutral DS. Thus, in the large $\rho_i/\lambda_D$ regime, the disappearance of the DS predicted by Stangeby is even more apparent.
}

In contrast, increasing $B_0$, and thus decreasing $\rho_i$, results in a progressive breaking of the above scaling (see Ref. \cite{ahedo97} for a discussion of the scales entering the transition).
For the case of low incidence angles, the consequences are twofold.
On the one hand, we observe a  stronger limitation of the ion flux perpendicular to the wall, as can be seen from Fig. \ref{fig:ion_vx_scanB0}. On the other hand, the charge separation near the wall tends to increase with $B_0$ (Fig. \ref{fig:rho_scanB0}).
These observations can be explained as follows. With increasing $B_0$, the CS extension becomes of the same order as that of the DS, so that the two sheaths overlap. Since the total potential drop remains constant, the overall width of the transition zone becomes too narrow to allow a quasi-neutral transition.
Consequently, the almost quasi-neutral transition previously observed for low magnetic fields at grazing incidence (curve corresponding to $\omega_{ci}/\omega_{pi} = 0.05$ in Fig. \ref{fig:rho_scanB0}) disappears, and the formation of a sheath is again required to ensure a smooth plasma-wall transition.
This effect may be interpreted as the appearance of a ``new" type of non-neutral sheath, whose thickness is of the order of the ion Larmor radius, when the scaling $\rho_i \approx \lambda_D$
is satisfied.


\subsection{Non-floating (biased) wall}
So far, we have considered stationary states for which the wall potential was left floating. We will now examine the effect of biasing the wall to a fixed
potential $\phi_{bias}^{W}$ below (ie, more negative than) the floating value $\phi_{float}^{W}$. Strictly speaking, the behaviour of the system in this case is not governed anymore by the ambipolarity condition at the wall, which was at the basis of the bounds obtained in Sect. \ref{sec:stangeby}. However, the ambipolarity condition can be reintroduced using the fact that the ion current density is the same in both situations, because it is fixed by the boundary condition at the CS entrance.

\begin{figure*}
\begin{center}
\subfigure[]{\includegraphics[width=0.45\textwidth]{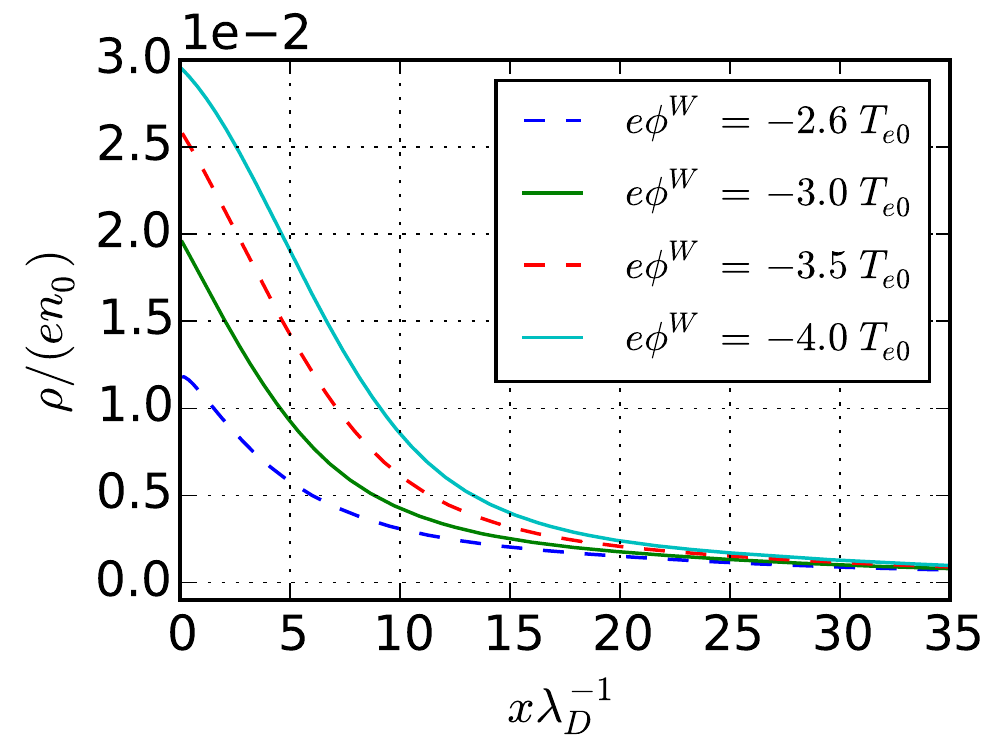}}
\subfigure[]{\includegraphics[width=0.45\textwidth]{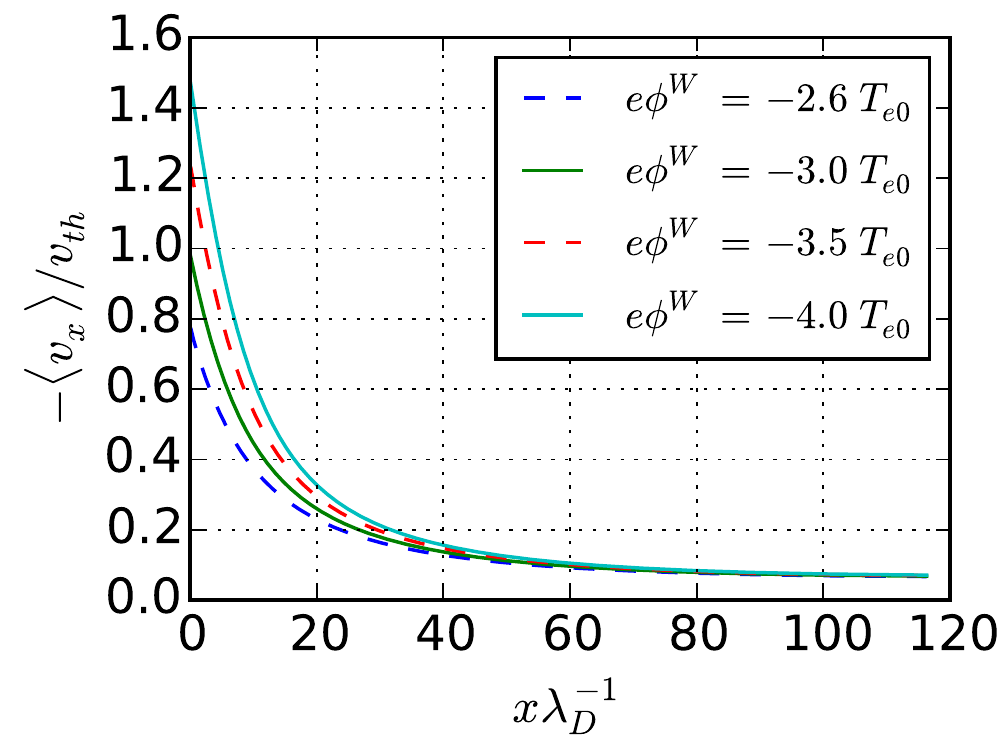}}
\caption{Collisionless case for $D^{+}$, $\alpha=2^{\circ}$, and $\omega_{ci}=0.05 \omega_{pi}$ with prescribed wall potential $\phi^{W}$ below the floating value $\phi_{float}^W \approx -2.5 T_{e0}$. (a) Charge density profiles; (b) Ion flow velocity perpendicular to the wall.}
\label{fig:rho_scan_phi_wall}
\end{center}
\end{figure*}

Still considering the electrons as perfectly magnetized up to the wall, we have
\begin{equation}
\begin{array}{c@{=}l}
\langle v_x \rangle_{bias} & \displaystyle \frac{\langle v_x \rangle_{float} n_{i,float}}{n_{i,bias}} \\
\ & \displaystyle \langle v_\parallel \rangle_e \sin \alpha \left(\frac{n_{e,bias}^W}{n_{e,bias}+\rho_{bias}/e}\right) \left(\frac{n_{e,float}^W}{n_{e,bias}^W} \right),
\end{array}
\end{equation}
leading to the modified bound
\begin{equation}
\vert \langle v_x \rangle_{bias} \vert = \sin \alpha \vert \langle v_\parallel \rangle_e^W \vert \exp \left [ \frac{e(\phi_{float}-\phi_{bias})}{T_{e0}}\right].
\end{equation}
Unsurprisingly, the bound on the ion flow velocity becomes less and less restrictive as the wall potential is set to lower values.  For a given target velocity, the corresponding critical angle decreases accordingly. Starting from a floating case, with a given (small) angle $\alpha$ for which the DS has nearly vanished, we can expect it to reappear as $\phi_{bias}$ is decreased. Considering for instance the deuterium case with $\alpha=2^{\circ}$, for which $e \phi_{float}^W  \approx -2.5 T_{e0}$, several biased-wall simulations were performed with different values of $\phi_{bias}^W$. An increase of the charge density near the wall is indeed observed (Fig. \ref{fig:rho_scan_phi_wall}a), resulting in the growth of the electric field (not shown here) and the ion flow velocity perpendicular to the wall (Fig. \ref{fig:rho_scan_phi_wall}b).

\begin{figure*}
\begin{center}
\subfigure[]{\includegraphics[width=0.45\textwidth]{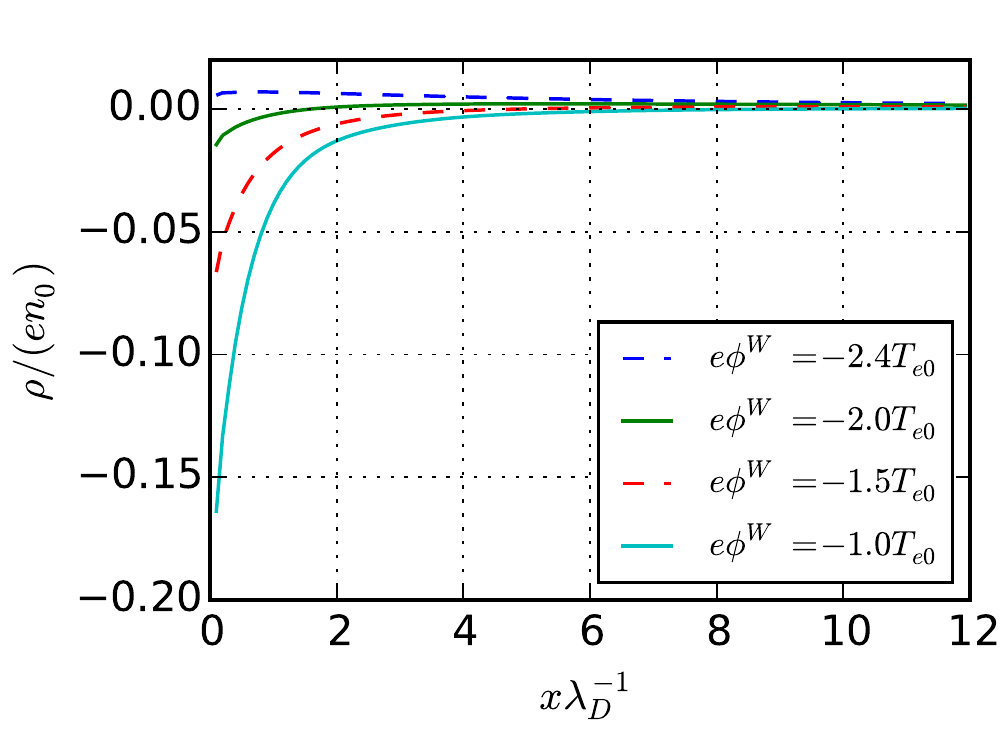}}
\subfigure[]{\includegraphics[width=0.45\textwidth]{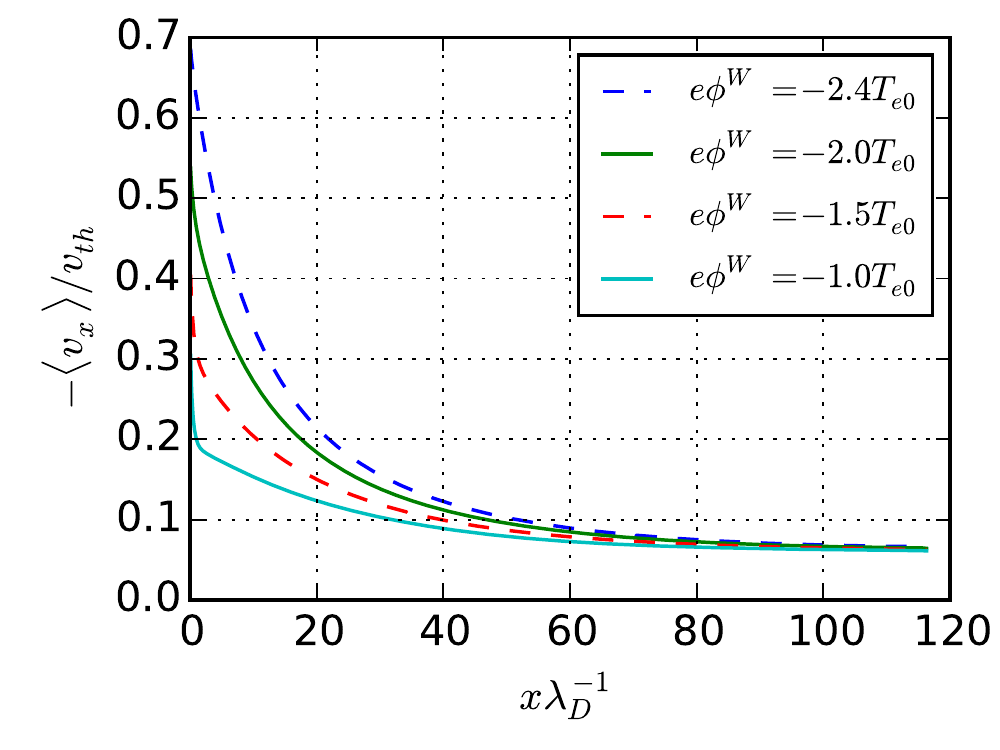}}
\caption{Collisionless case for $D^{+}$, $\alpha=2^{\circ}$, and $\omega_{ci}=0.05 \omega_{pi}$ with prescribed wall potential $\phi^{W}$ {\em above} the floating value $\phi_{float}^W \approx -2.5 T_{e0}$. (a) Charge density profiles; (b) Ion flow velocity perpendicular to the wall.}
\label{fig:rho_scan_phi_wall_pos}
\end{center}
\end{figure*}

{\modif
We also analysed the case of a wall biased at a potential above (ie, less negative than) the floating value, a situation relevant to tokamak divertors where the divertor tiles may be biased positively with respect to the floating potential. The results are depicted in Fig. \ref{fig:rho_scan_phi_wall_pos}. For grazing incidence ($\alpha = 2^\circ$), a small bias above the floating potential is sufficient to remove completely the charge separation near the wall or even to reverse its sign. At the same time, the ion velocity at the DS entrance drops well below the sound speed. The conclusion here is that, for grazing incidence, a small bias above the floating potential can remove any remnants of the DS, so that the transition to the wall is truly charge-neutral and subsonic. For almost normal incidence, the necessary bias would have been much larger.
}

\section{Comparison between the fluid model and kinetic simulations}
\label{sec:fluid_kin_comp}

The results of Stangeby \cite{Stangeby2012} were obtained using simple fluid model that had been proposed earlier by Chodura \cite{Chodura82} and Riemann \cite{Riemann94}, {\modif and further developed in \cite{Gao2003}}. Although its predictions are basically correct, most notably the disappearance of the DS for low incidence angles, it would be interesting to test its limitations by comparing the fluid results to those of our kinetic code.

Taking the velocity moments of Eq. \eqref{eq:vlasov} up to first order yields the following fluid system in the stationary
state
\begin{subequations}
\begin{align}
\partial_x(n_i u_x) &= -\nu_i (n_i-n_{i0}) \\
u_x \partial_x u_x &= -\frac{\partial_x P_{xx}}{m_i n_i} - \frac{q_i}{m_i} \partial_x \phi -\omega_y  u_z   -\nu_i \frac{n_{i0}}{n_i} u_x \\
u_x \partial_x u_y &= \left\lbrace -\frac{\partial_x P_{xy}}{m_i n_i} \right\rbrace + \omega_x  u_z -\nu_i \frac{n_{i0}}{n_i} u_y \\
u_x \partial_x u_z &=  \left\lbrace -\frac{\partial_x P_{xz}}{m_i n_i} \right\rbrace + \omega_y u_x - \omega_x u_y  -\nu_i \frac{n_{i0}}{n_i} u_z,
\end{align}
\label{eq:fluid_system}
\end{subequations}
where $u_k=\langle v_k \rangle, k=x,y,z$, $\omega_x =\omega_{ci} \sin \alpha$,  $\omega_y=\omega_{ci} \cos \alpha$, {\modif and $n_{i0}$ is the bulk density}. In the Chodura-Riemann-Stangeby (CRS) model for the collisionless magnetic presheath, we have $\nu_i=0$ and  two assumptions are made:  (i) the non-diagonal components of the kinetic pressure tensor [terms in braces in Eqs. \eqref{eq:fluid_system}c-d] are neglected and (ii) the $xx$ component of the pressure tensor is assumed to follow an isothermal closure
$P_{xx} = T_{0} n_i$, with constant $T_{0}$.
Combined with the quasi-neutrality relation and the Boltzmann law for the electron density, the system of Eqs. \eqref{eq:fluid_system}a-d can be integrated easily \cite{Chodura82,Stangeby2012}. In Ref. \cite{Stangeby2012}, the system is integrated in $x$ starting from the CS exit. In our case, as the kinetic simulation encompasses both the CS and the DS, defining the CS exit would require setting a somewhat arbitrary threshold on the charge separation. Thus, in order to compare our simulation results with the CRS fluid model, we integrate the fluid equations starting from the CS entrance at $x=L$.
{\modif
In order to compare with the kinetic results, the constant temperature $T_0$ of the fluid model is set equal to the value of $T_{xx} \equiv P_{xx}/n$ obtained from the ion distribution $f^{in}$ at the CS entrance, given in Eq. (\ref{eq:fin})
\footnote{\modif In our case, we have $T_0=T_{i0}(1-0.55 \sin^2 \alpha)$, where $T_{i0}=T_{e0}$ is the parameter appearing in Eq. (\ref{eq:fin}).}. For clarity, as our notation differs from that used in Ref. \cite{Stangeby2012}, the explicit form of the CRS fluid equations is given in Appendix \ref{appendix:fluid_model}.
}

In Fig.  \ref{fig:ux_chodura} we compare the average velocity $\langle v_x \rangle$ extracted from the simulation data with $u_x$ computed from the fluid model for a few values of $\alpha$. While the agreement is quite good for $\alpha = 3^{\circ}$ and $5^{\circ}$, discrepancies appear for larger angles. It is important to note that those discrepancies arise before charge separation becomes significant, ie, when the plasma can still be reasonably considered as quasi-neutral ($x > 10\lambda_D$). Proceeding to the same comparison for the $y$ and $z$ components of the average velocity (Figs. \ref{fig:uy_uz_chodura}a and \ref{fig:uy_uz_chodura}b), we observe quite similar discrepancies on $u_z$ but far larger and systematic ones for $u_y$ on nearly the whole domain.
Thus, as far as only the $u_x$ profiles are concerned (and consequently also the potential profiles), the predictions of the fluid model in the CS can be considered as rather good for the lowest range of incidence angles. The somewhat large and systematic discrepancies observed for the other velocity components would require closer scrutiny. They probably arise from the violation of both assumptions made in the fluid model.

\begin{figure}
\begin{center}
\includegraphics[width=0.45\textwidth]{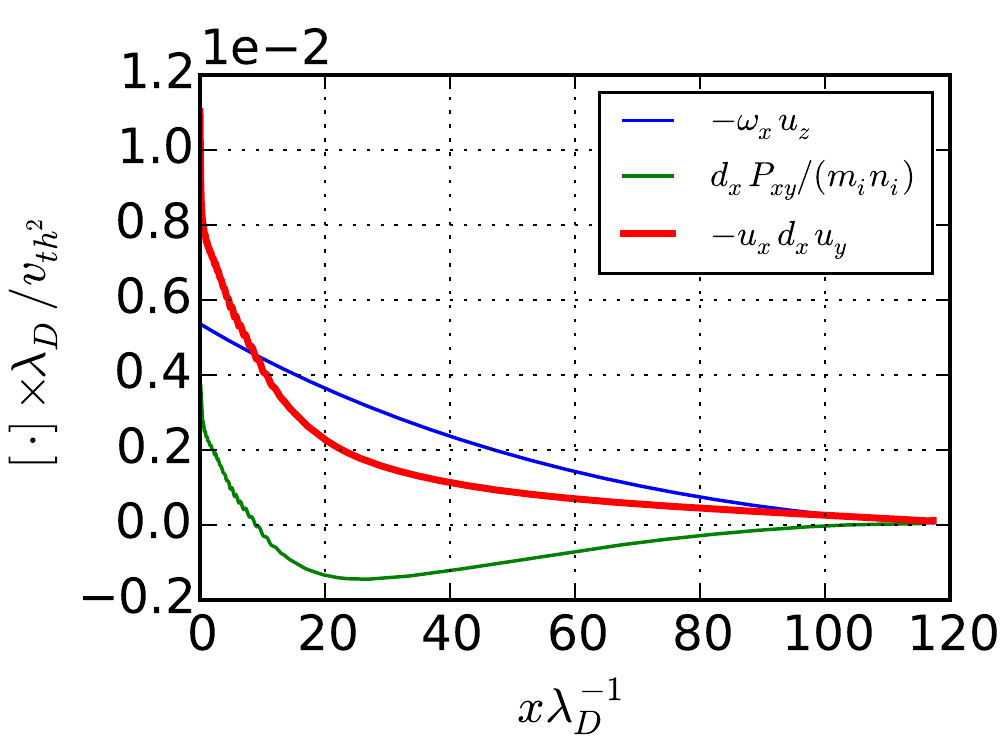}
\caption{Various terms of the momentum balance equation along $y$ [Eq. \eqref{eq:fluid_system}c] computed from the collisionless kinetic simulations, for $\alpha=3^{\circ}$ and deuterium ions.}
\label{fig:fluid_balance_D_alpha3_uy}
\end{center}
\end{figure}

To refine our comparison, we computed, from the kinetic simulations, the various terms entering the $y$ and $z$ components of the momentum balance equations \eqref{eq:fluid_system}c-d.
The comparison indicates that the contribution of the non-diagonal terms of the pressure tensor is not negligible.
Focusing in particular on the equation for $u_y$, Fig. \ref{fig:fluid_balance_D_alpha3_uy} shows that the term containing $P_{xy}$ is comparable to the other terms, even in the CS.
In contrast, the term $P_{xz}$ (not shown here) is indeed negligible.
We emphasize the fact that the non-diagonal nature of the pressure tensor is not an artifact due to the choice of coordinates, which could be eliminated by using a field-aligned basis: although the distribution at the CS entrance is indeed separable in $(v_\parallel,{\mathbf v}_\perp)$, this separability is lost during the transition.

\begin{figure}
\begin{center}
\includegraphics[width=0.45\textwidth]{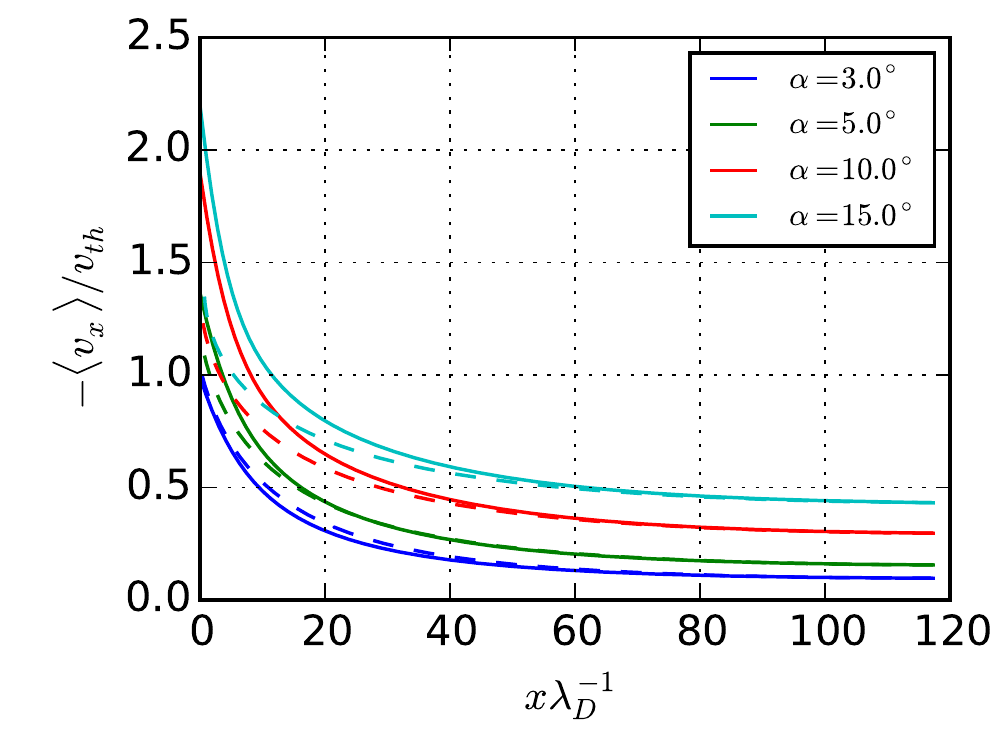}
\caption{Ion mean flow perpendicular to the wall from the collisionless kinetic simulations (continuous lines) and the CRS model (dashed lines), for various values of $\alpha$, and deuterium ions. {\modif Note that the CRS model, being quasi-neutral, is not meaningful in the DS.}}
\label{fig:ux_chodura}
\end{center}
\end{figure}
\begin{figure*}
\begin{center}
\subfigure[]{\includegraphics[width=0.45\textwidth]{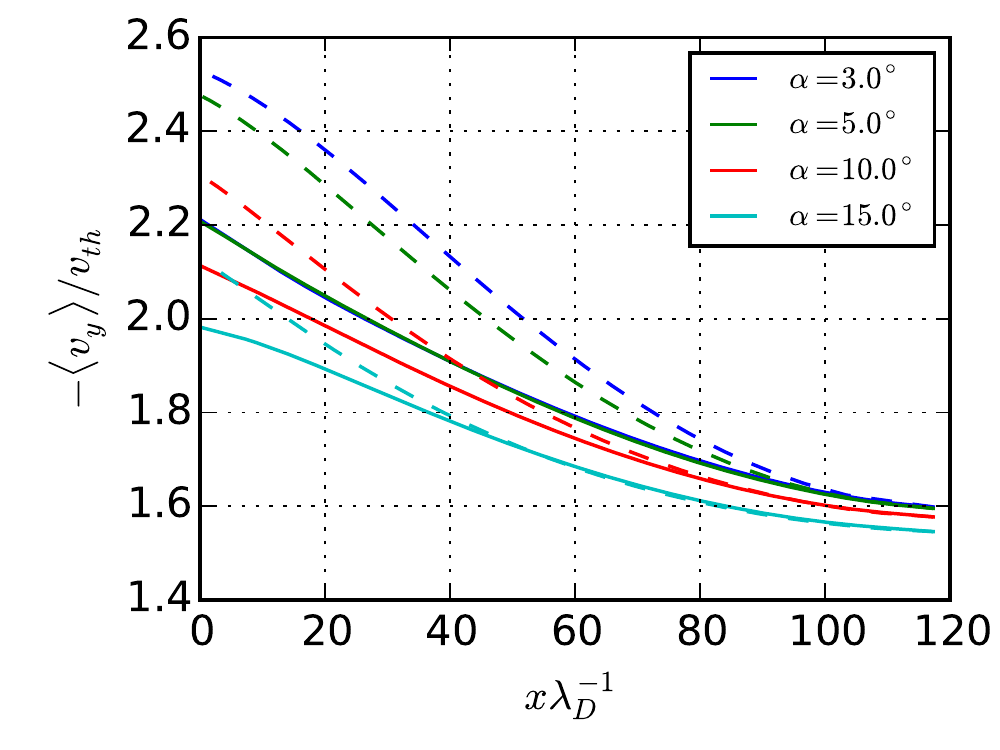}}
\subfigure[]{\includegraphics[width=0.45\textwidth]{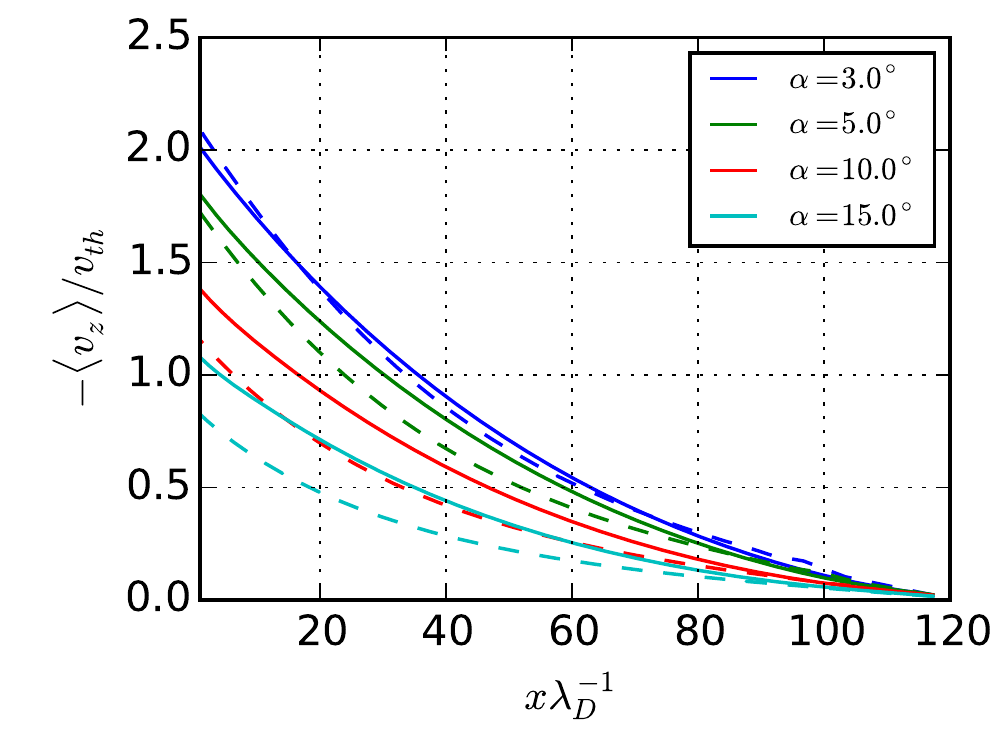}}
\caption{Same as Fig. \ref{fig:ux_chodura} for $u_y$ (a) and $u_z$ (b). {\modif Note that, in the left panel (a), the continuous curves for $\alpha=3^\circ$ and $5^\circ$ are virtually indistinguishable.}}
\label{fig:uy_uz_chodura}
\end{center}
\end{figure*}

\begin{figure}
\begin{center}
\includegraphics[width=0.45\textwidth]{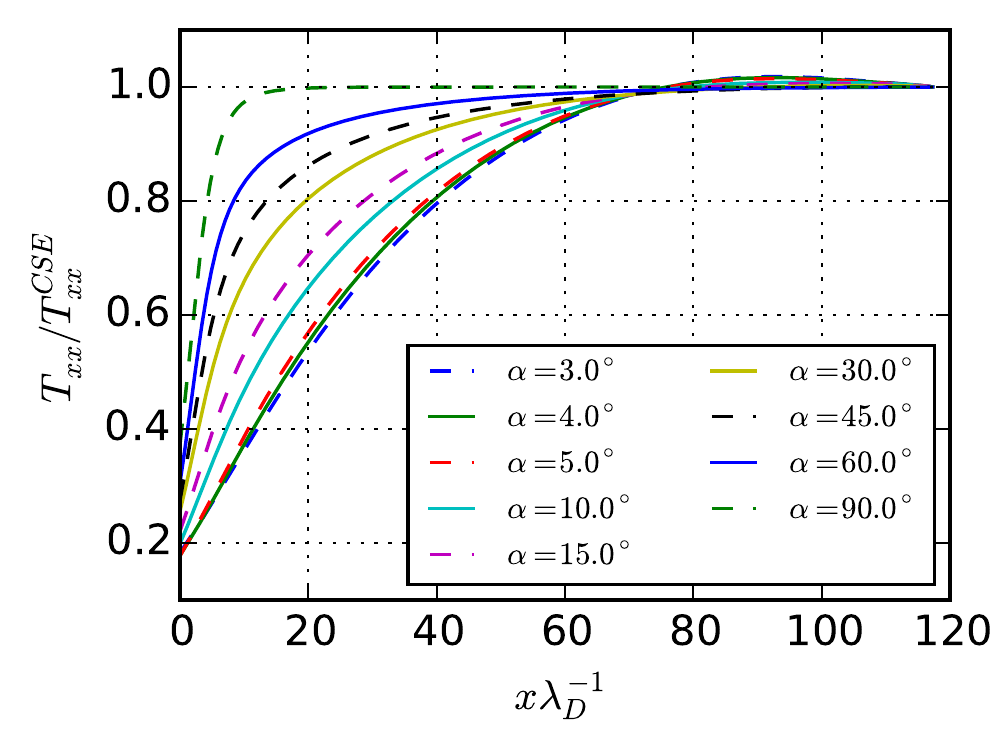}
\caption{Spatial variation of the temperature $T_{xx}$ normalized to its value at the CS entrance $T_{xx}^{CSE}$, for collisionless simulations with deuterium ions.}
\label{fig:T_xx_D_nocoll}
\end{center}
\end{figure}

Let us now consider the validity of the isothermal closure for the $P_{xx}$ component of the pressure tensor. In the normal incidence case, for which only the DS exists, the temperature $T_{xx}$ (ie, the variance along $v_x$) decreases as the ion population is accelerated towards the wall by the electric field. This well-known "acceleration cooling" \cite{Anton1978,coulette2015} persists in the magnetized case. More importantly, as the electric field profile is spread out with decreasing $\alpha$, $T_{xx}$ has a non-negligible variation over both the DS and CS. This is clearly visible in Fig. \ref{fig:T_xx_D_nocoll} showing the evolution of $T_{xx}$ relative to its value at the CS entrance (we recall here that $T_{xx}^{CSE}$ depends on $\alpha$, see Sect. \ref{sec:bc_collisionless}). As a consequence, though the isothermal closure may be considered a reasonable approximation (outside the DS) for the large-to-intermediate angle range, it becomes clearly invalid in a large part of the transition layer for smaller angles of incidence.

\begin{figure*}
\begin{center}
\subfigure[]{\includegraphics[width=0.45\textwidth]{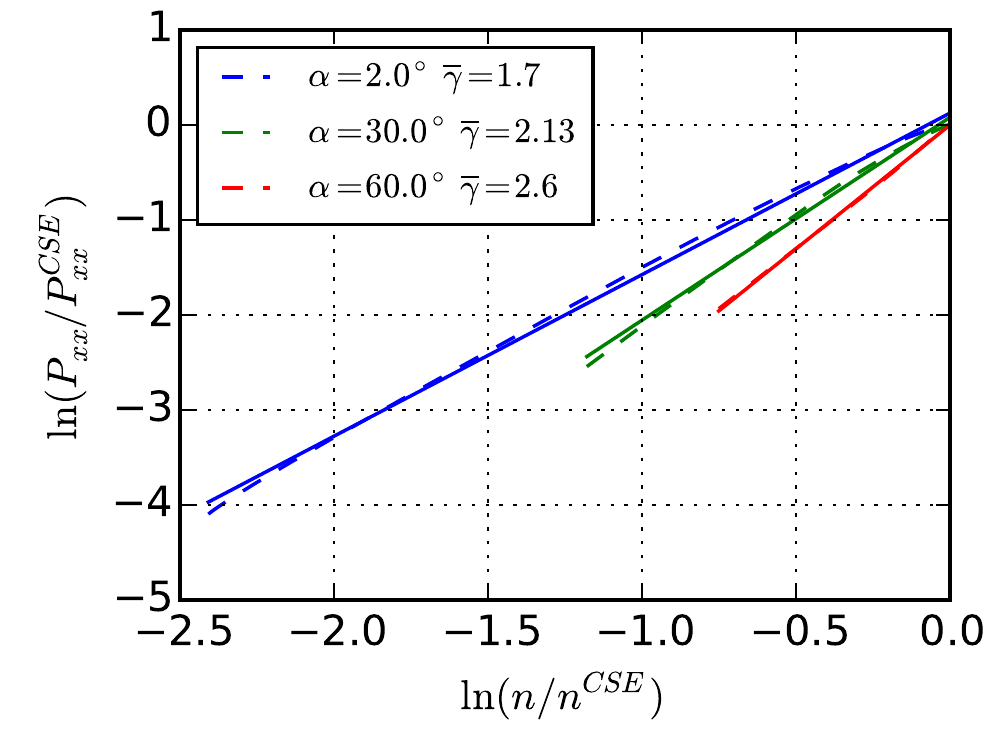}}
\subfigure[]{\includegraphics[width=0.45\textwidth]{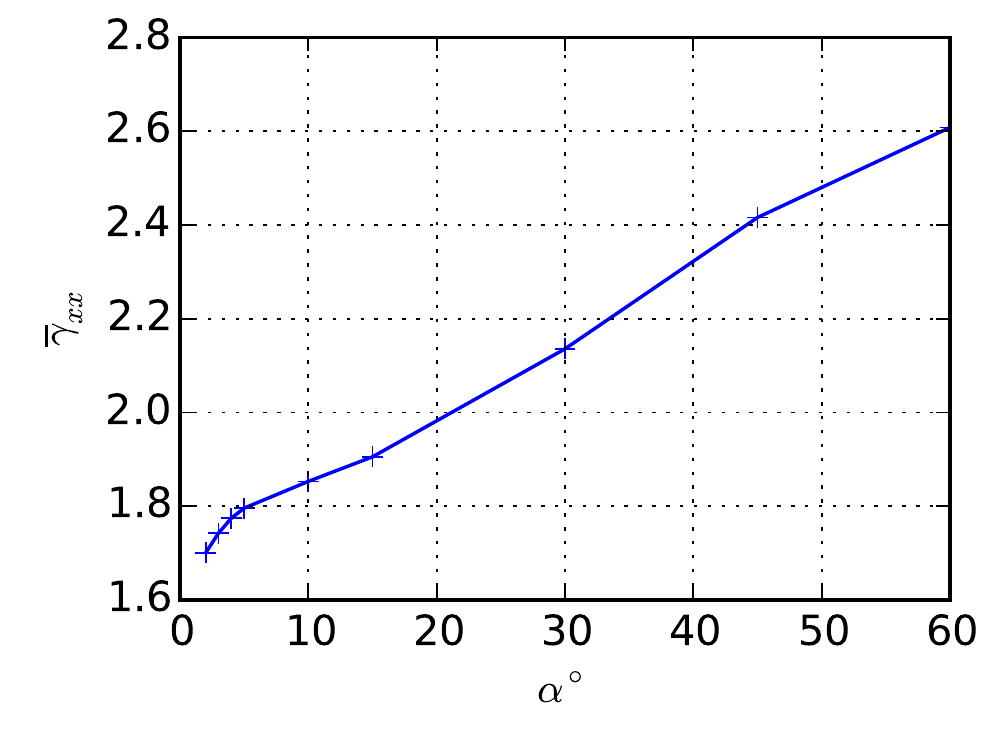}}
\caption{(a) Determination of an average polytropic index $\overline{\gamma}$ by linear regression of $\ln P_{xx}=f(\ln (n))$ for $\alpha \in \lbrace 2^{\circ},30^{\circ},60^{\circ}\rbrace$. Simulation data are plotted as dashed lines and regression results as continuous lines;
(b) Behaviour of the average index $\overline{\gamma}$ with $\alpha$.}
\label{fig:gamma_xx_vs_alpha_D_nocoll}
\end{center}
\end{figure*}

Having established that the isothermal closure does not fit the actual behavior of the distribution for low $\alpha$, one may hope to fit a slightly more general polytropic closure $d(\ln P_{xx})= \gamma d(\ln n)$. A typical constant polytropic coefficient $\overline{\gamma}$ can be obtained by linear regression for each value of $\alpha$ (Fig. \ref{fig:gamma_xx_vs_alpha_D_nocoll}). We observe a large variation with $\alpha$, as can be expected when going from the two-scale behaviour at large $\alpha$ to the smoother transition at low $\alpha$ (see Fig. \ref{fig:T_xx_D_nocoll}).
{\modif
For $\alpha=90^\circ$, the CS disappears altogether and the quasi-neutral fluid model cannot be meaningfully compared to the kinetic results.
}
Alternatively, one could compute a local polytropic coefficient $\gamma(x)=d(\ln P)/d(\ln n)$ \cite{Tskhakaya2005}, but this yields very large variations over the domain and with $\alpha$, and is prone to numerical instability in the low-gradient zones.

\begin{figure}
\begin{center}
\includegraphics[width=0.45\textwidth]{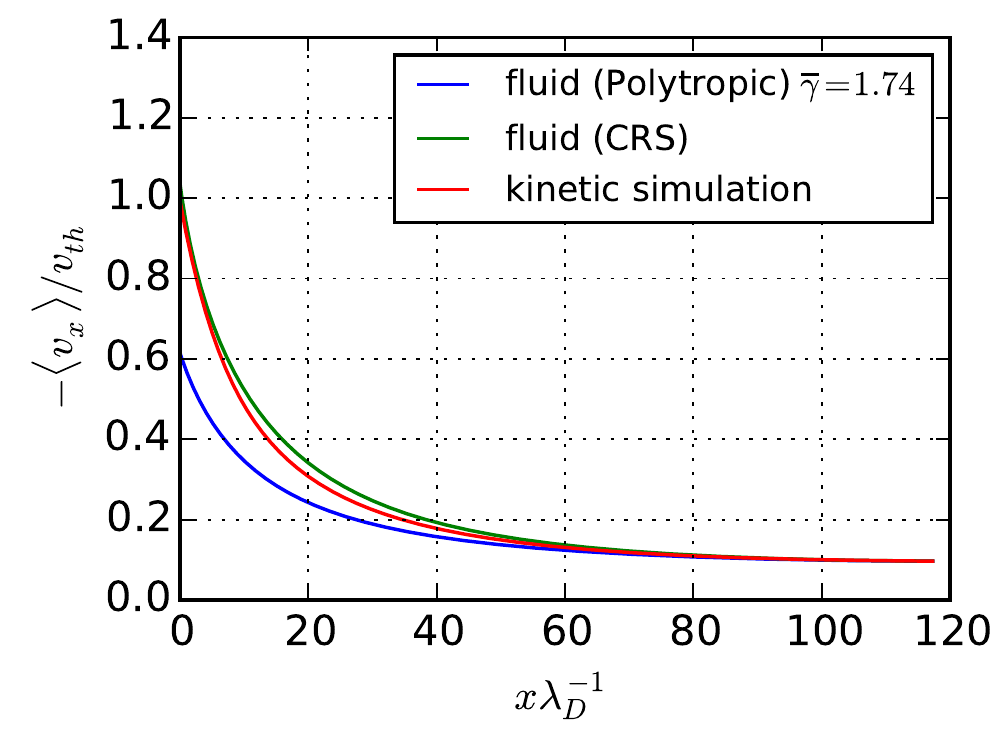}
\caption{Ion mean velocity profile $\langle v_x \rangle$ for the kinetic simulation, the fluid CRS isothermal model ($\gamma=1$), and the fluid polytropic model ($\gamma=1.74$). Deuterium ions with incidence angle $\alpha=3^\circ$.}
\label{fig:ux_comparison}
\end{center}
\end{figure}

{\modif
We also tried to use the computed exponent $\gamma$ to improve the match between the kinetic and the fluid models (using, in the latter, a polytropic equation of state, $P_{xx}\propto n_i^{\gamma}$), but this does not seem to work well for $u_x$ (Fig. \ref{fig:ux_comparison}). The profiles of the mean velocities along $y$ and $z$ are not improved either, which is not surprising as their discrepancy with the kinetic data comes primarily from the assumption of isotropic pressure.}
The main conclusion here is that it is not possible to match the kinetic simulation data with a simple polytropic closure.

All in all, the comparison between the simulation results and the predictions of the fluid model leads us to conclude that: (i) a rather good agreement is obtained for the $u_x$ profile (and consequently for the potential profile) for the lowest values of $\alpha$, but (ii) a worse agreement is observed for the other components of the mean velocity, due to the violation of some of the assumptions of the fluid model.


\section{Collisional simulations}
\label{sec:coll}
In the preceding collisionless simulations, the field-aligned ion flow velocity at the CS entrance was imposed through an ad-hoc boundary condition. In order to ensure that such results are not specific to the collisionless regime, we performed a series of collisional simulations. In this case, the simulation domain is much larger (typically $2\times 10^4 \lambda_D$) in order to encompass the full transition {\modif from the the plasma bulk to the wall. The ion velocity distribution in the bulk is an isotropic Maxwellian with temperature $T_{i0}=T_{e0}$.}
Then, the distribution function at the CS entrance is no longer imposed as a boundary condition, but rather arises self-consistently in the collisional presheath located upstream the DS.
A thorough characterization of the transition, using the same kinetic model, was performed by Devaux et al. \cite{devaux2006}. Here, we will focus on the question whether collisions modify the results obtained in the collisionless regime for grazing incidence.

{\modif As in the collisionless simulations, parametric scans} in $\alpha$ were performed for the same range of angles for $\omega_{ci}=0.05 \omega_{pi}$, with three values of the collision frequency $\nu_i \in \left\lbrace 5\times 10^{-4},10^{-3}, 5 \times 10^{-3}\right \rbrace v_{th} \lambda_D^{-1}$. For this range of parameters the transition is characterized by the scaling $\lambda_D \ll \rho_i \ll \lambda_{mfp}$, where $\lambda_{mfp}=v_{th}/\nu_i$. This is the intermediate ${\bf B}_0$ regime described in {\modif Refs. \cite{ahedo97, devaux2006, Kos2014}}, for which the collisional presheath, Chodura sheath, and Debye sheath are well separated.

\begin{figure}
\begin{center}
\includegraphics[width=0.5\textwidth]{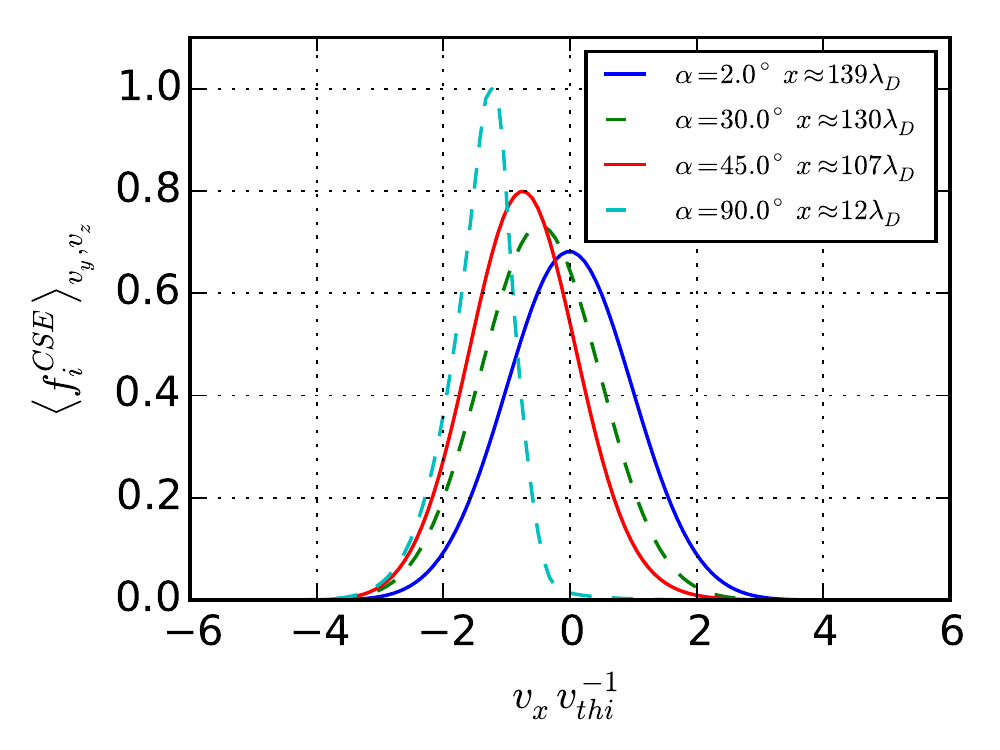}
\caption{Ion velocity distribution functions at the CS entrance, for various angles of incidence, $\nu_i=5\times 10^{-4}v_{th} \lambda_D^{-1}$ and deuterium ions. The position of the CS entrance is indicated in the inset. In order to facilitate the comparison, each distribution function is normalized in such a way that it has the same maximum as the prescribed collisionless distribution of same incidence angle (Fig. \ref{fig:f_in_alpha_phi_wall_nocoll}a).}
\label{fig:fin_coll}
\end{center}
\end{figure}

{\modif
As a preliminary benchmark, we use the collisional simulations to check the validity of the boundary condition that we prescribed at the entrance of the CS in the collisionless runs [Eq. (\ref{eq:fin})]. For this, we need a criterion to define the CS entrance. In the collisional presheath, the mean ion velocity is aligned with the magnetic field: the CS entrance corresponds to the location where this alignment breaks down. As a quantitative criterion, we took a deviation of $0.25^\circ$ with respect to the angle of incidence $\alpha$. The computed distribution functions are shown in Fig. \ref{fig:fin_coll} and look very much like the prescribed distributions used in the collisionless runs (Fig. \ref{fig:f_in_alpha_phi_wall_nocoll}a).
}

\begin{figure*}
\begin{center}
\subfigure[]{\includegraphics[width=0.45\textwidth]{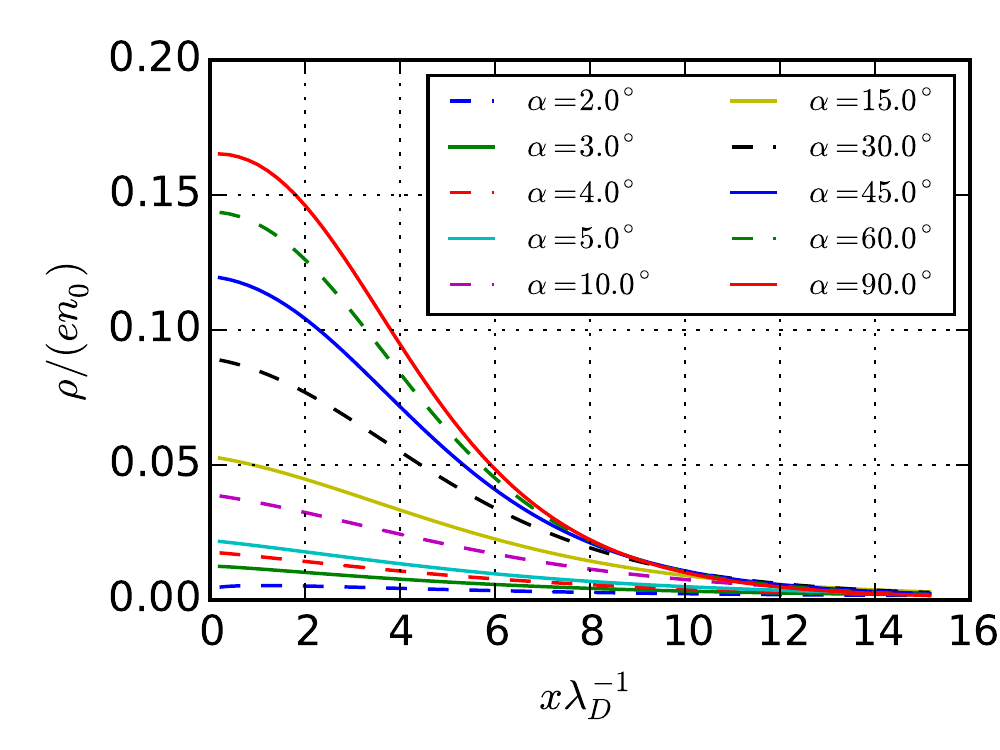}}
\subfigure[]{\includegraphics[width=0.45\textwidth]{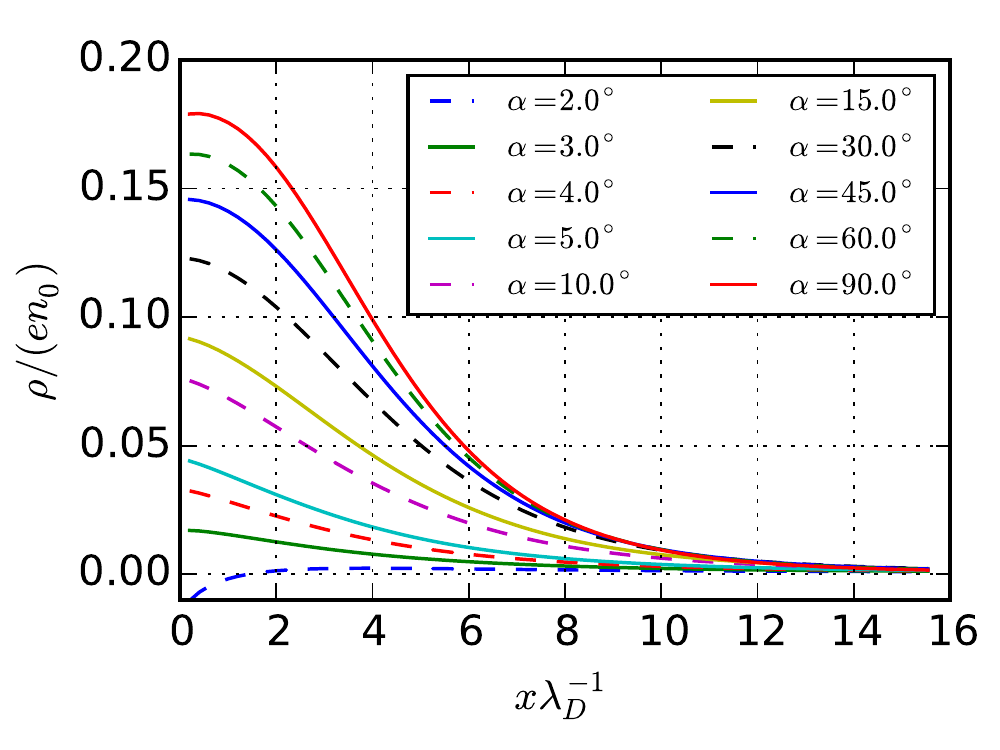}}
\caption{Charge density profiles near the wall for two collisional simulations with deuterium ions. The collision frequencies are $\nu_i = 5 \times 10^{-4} v_{th}/\lambda_D$ (a) and $\nu_i = 5 \times 10^{-3} v_{th}/\lambda_D$ (b).}
\label{fig:rho_coll}
\end{center}
\end{figure*}

We can now verify the robustness of Stangeby's result in the collisional regime.
First and foremost, we still observe a decrease of the charge density near the wall for decreasing angles of incidence (Fig. \ref{fig:rho_coll}), with similar consequences on the electric field and potential profiles near the wall (not shown). The principal effect analysed in this work is thus not destroyed by the presence of collisions.

\begin{figure}
\begin{center}
\includegraphics[width=0.45\textwidth]{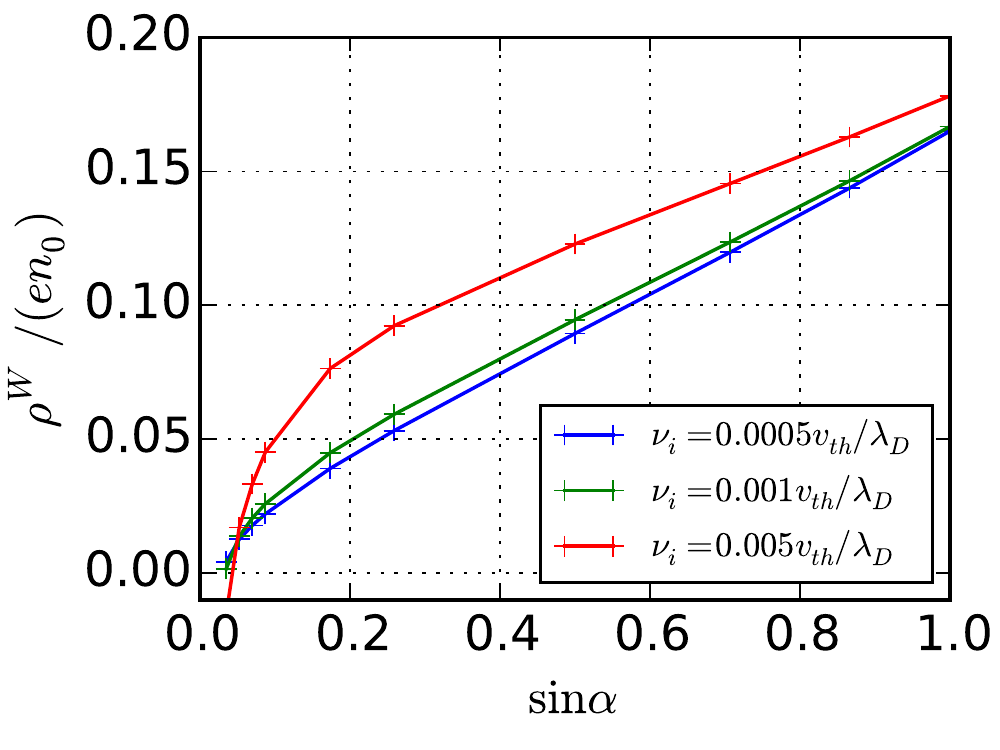}
\caption{Charge density on the wall as a function of the incidence angle $\alpha$, for three values of the collision frequency $\nu_i$. Deuterium ions.}
\label{fig:rho_w_alpha}
\end{center}
\end{figure}

\begin{figure}
\begin{center}
\includegraphics[width=0.45\textwidth]{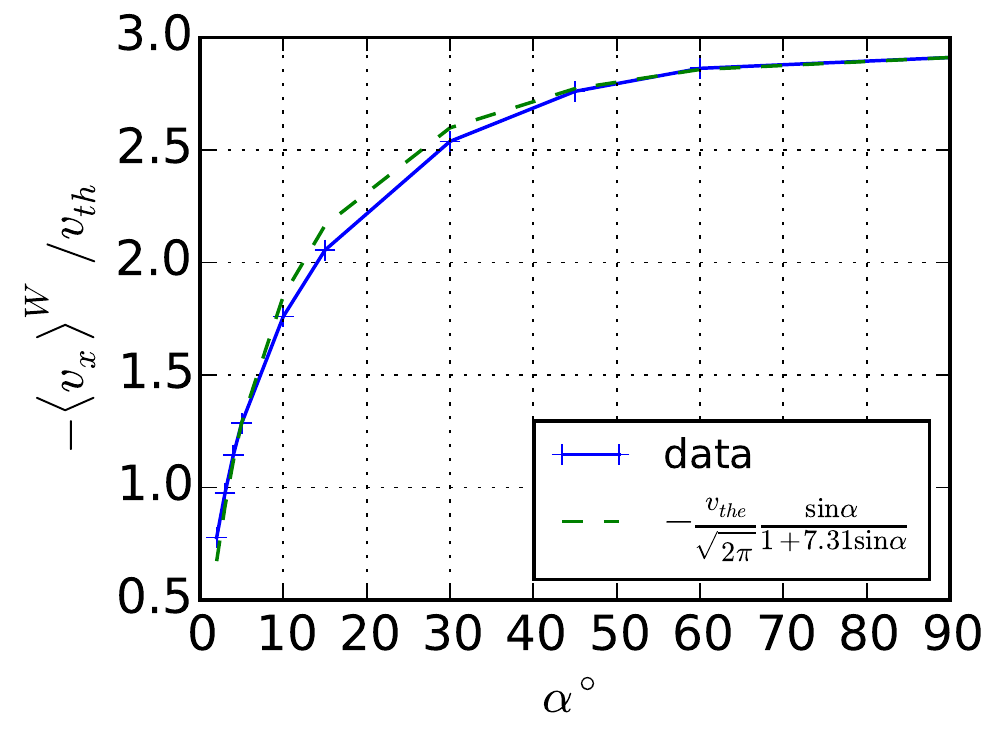}
\caption{Ion flow velocity on the wall as a function of the incidence angle $\alpha$, for a collisional case with frequency $\nu_i = 5 \times 10^{-4} v_{th}/\lambda_D$.
The numerical coefficient $\kappa=7.31$ of the semi-empirical law Eq. \eqref{eq:vx_wall_fit} is computed by exact interpolation from the case $\alpha=90^{\circ}$.  Deuterium ions.}
\label{fig:vx_wall_vs_alpha_coll5em4_ref90}
\end{center}
\end{figure}

Second, the nearly linear dependency of the wall charge density with $\sin\alpha$
(which was observed in the collisionless case, see Fig. \ref{fig:rho_vx_wall_vs_alpha_nocoll}a)
is slightly perturbed by the collision terms as shown in Fig. \ref{fig:rho_w_alpha} {\modif(note that here the charge density is normalized to the value $n_0$ in the bulk plasma, whereas in the preceding sections the normalization value $n_0$ referred to the density at the CS entrance)}. A marginal sign inversion of $\rho$ near the wall can even be observed in the $(\alpha=2^{\circ},  \nu_i= 5 \times 10^{-3} v_{th}\lambda_D^{-1}$) case. Despite this perturbation, the ion perpendicular flow as a function of $\alpha$  may still be roughly fitted by the same semi-empirical law as in the collisionless case (Fig. \ref{fig:vx_wall_vs_alpha_coll5em4_ref90}).

\begin{figure}
\begin{center}
\includegraphics[width=0.45\textwidth]{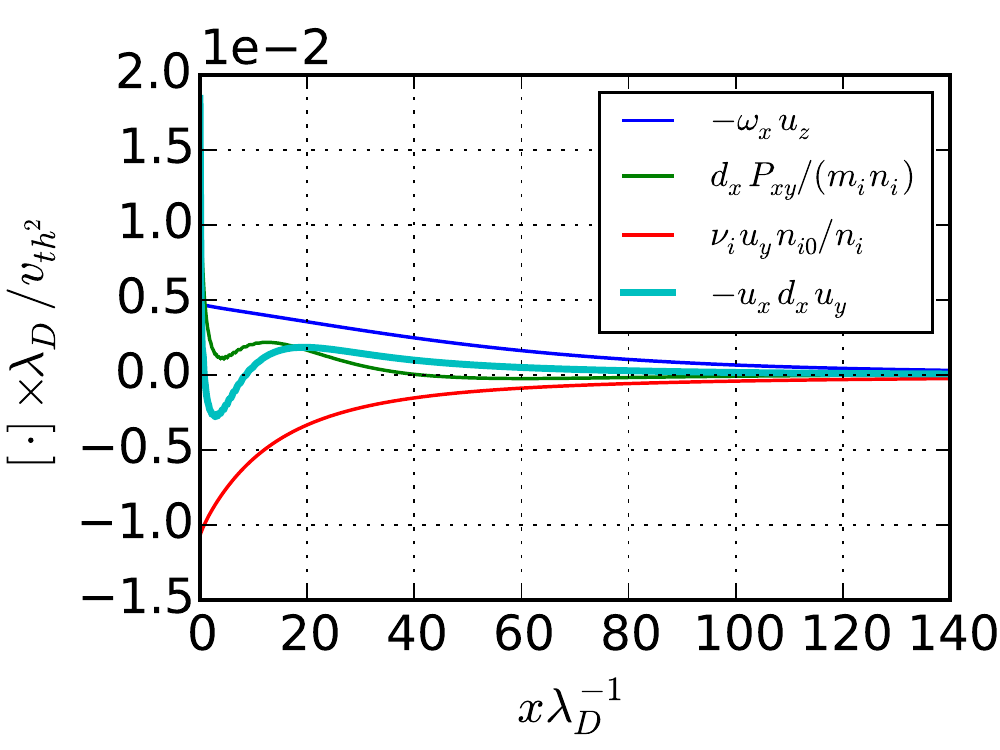}
\caption{Various terms in the momentum balance equation along $y$, Eq. \eqref{eq:fluid_system}c. Collisional simulation with $D^{+}$ ions, $\alpha=3^{\circ}$, and $\nu_i = 5 \times 10^{-4} v_{th}/\lambda_D$. }
\label{fig:fluid_y_D_5em4}
\end{center}
\end{figure}

Last, let us extend the analysis of the various terms entering the fluid momentum balance in Eqs. \eqref{eq:fluid_system}b-d. Setting aside the additional impact of the friction terms specific to our collision model, we still observe a non-negligible impact of the non-diagonal term of the pressure tensor $P_{xy}$ in the fluid momentum balance along the $y$ axis (Fig. \ref{fig:fluid_y_D_5em4}). As was the case for the collisionless regime, the $P_{xz}$ cross-term (not shown here) is indeed small outside the space-charge region near the wall.

\section{Conclusions and pending issues}
\label{sec:conclusions}

The main focus of this paper was on the observation, made by Stangeby \cite{Stangeby2012}, that the Debye sheath should disappear when the plasma is immersed in a magnetic field with grazing angle of incidence with respect to the wall. Stangeby's result was deduced from a simple 1D fluid model with Boltzmann electrons and isothermal closure for the ions. Thus, it was worth to check whether the result holds under less stringent conditions on the ion model, namely using a kinetic rather than fluid approach.

Our calculations showed clearly that the main result holds: the charge separation progressively disappears for smaller and smaller angles of incidence, and the ion flow velocity perpendicular to the wall is limited to subsonic speeds.
Though no critical angle arises due to the lack of singularity at the DS entrance in the kinetic model, the overall behaviour is consistent with the predictions of Ref.  \cite{Stangeby2012}.  We also confirmed the increased spreading, with decreasing $\alpha$, of the electric field and plasma density over distances of several Larmor radii from the wall. These features appear in both collisionless and collisional simulations, and may thus be considered as robust, provided the scaling $\lambda_{coll} \gg \rho_i \gg \lambda_D$ is satisfied.

As noted by Stangeby \cite{Stangeby2012}, the spreading of the electric field and plasma density further from the wall (compared to what is usually expected from simpler models) has important consequences on the recycling of sputtered particles in a tokamak edge. It should be taken into account, whenever possible, in the computational codes that deal with plasma edge recycling.

Further, by comparing the kinetic and fluid profiles, we found that, although a rather good quantitative agreement on the ion flow velocity perpendicular to the wall can be obtained for small angles, the assumptions of a scalar pressure tensor and isothermal closure in the fluid model are clearly violated. These findings point at the limitation of the fluid models usually employed to study this type of scenarios.

Finally, in all simulations apart from the most collisional ones, we observed a rather robust linear scaling of the charge density at the wall with $\sin \alpha$. As a consequence, the value of the ion mean flow {\modif velocity} perpendicular to the wall obeys the simple semi-empirical law: $\langle v_x \rangle_i^W= v_{the}/\sqrt{2\pi} \sin \alpha/(1+\kappa \sin \alpha)$, where $\kappa$ is a coefficient that can be determined from a single simulation at normal incidence.

All the previous considerations are correct as far as the various simplifying assumptions made both in the fluid and kinetic models are satisfied. The first concerns the electrons, which were assumed to be perfectly magnetized up to the wall and to follow a Boltzmann law. For very small angles of incidence ($\alpha < 1^{\circ}$), these assumptions cease to be valid and the electron dynamics should be treated with a fully kinetic model.

A second assumption lies in the reduction of the system to one dimension in space. For divertor targets, the determination of the CS and DS structure near the inter-tile gaps would require at the very least a two-dimensional model in space, encompassing the full incidence plane of the magnetic field [ie, the plane $(x,y)$ in our geometry, see Fig.  \ref{fig:geometry}] in order to properly determine both the structure of the electric field and the particle flows in those regions.
Of course, an extra spatial dimension would increase dramatically the complexity of the present kinetic code. Nevertheless, it is an important feature that needs to be addressed for quantitative comparisons with tokamak measurements.

\section*{Acknowledgments}
We thank Stéphane Heureaux for several useful discussions.
The authors acknowledge the support of the French Agence Nationale de la Recherche (ANR), project PEPPSI, reference ANR-12-BS09-028-01.

\appendix
\section{Collisionless fluid model}
\label{appendix:fluid_model}

The following relations are established from the fluid system \eqref{eq:fluid_system} in the collisionless case ($\nu_i=0$) using a diagonal pressure tensor ($P_{xy}=P_{xz}=0$) and an isothermal closure $P = n T_0$. Exact neutrality $n_i=n_e$ and a Boltzmann law for electrons are assumed. The integration of the system follows the same pattern as in Refs. \cite{Chodura82,Stangeby2012}, the only difference being in the fact that no assumptions were made on the value of the boundary conditions (ie, they are a priori unrelated to $c_s$).

Starting from a reference point $x_0$ with fluid velocities $(u_{x0}<0,u_{y0},u_{z0})$, the position $x_1<x_0$ where $u_x$ reaches the value $u_{x1}$  is obtained through the integral expression:
\begin{equation}
\omega_{ci} \cos\alpha (x_1-x_0)= - \int\limits_{u_{x0}}^{u_{x1}}
\frac{ u\left(1- \frac{c_s^2}{u^2}\right) du}
{
[D(u)]^{\frac{1}{2}}
},
\label{eqapp:master_int_eq}
\end{equation}
with
\begin{equation}
D(u)=U_0^2 +c_s^2 \ln \left(\frac{u}{u_{x0}}\right)^2-u^2-u_y(u)^2,
\end{equation}
\begin{equation}
u_y(u)= u_{y0}  -\tan \alpha \left[ (u-u_{x0}) + c_s^2\left(\frac{1}{u}-\frac{1}{u_{x0}}\right)
\right],
\label{eq:fluid_model_uy}
\end{equation}
and $U_0^2= u_{x0}^2+u_{y0}^2+u_{y0}^2$, $c_s^2=(T_0+T_{e0})/m_i$. The above relations are obviously valid only as long as $D(u)$ does not vanish in the integration range.  Here, the $c_s$ factor arises solely from the isothermal closure for the ions, and does not depend on the boundary conditions.

From a numerical point of view, the velocity profile $u_x(x)$ is reconstructed as follows: a uniform discrete velocity grid $(u_n,n=0\dots N)$ is generated between $u_0=u_x^{CSE}$ and $u_N= \max(-\vert u_{sing} \vert,-u_{bound})$, where $u_{sing}$ is the singular velocity cancelling $D(u)$ and $u_{bound}$ is the velocity bound obtained from Eq. \eqref{eq:cs_bound_ion}. Starting from $[u_0,u_1]$ Eq. \eqref{eqapp:master_int_eq} is integrated over each pair $[u_n,u_{n+1}]$. The end result is a sequence $[x_0,\dots,x_N]$ of positions matching the velocities $[u_0,\dots,u_N]$. The $u_y$ profile is obtained directly using Eq. \eqref{eq:fluid_model_uy}.
The velocity $u_z$ is recovered from $u_x$ using
\begin{equation}
u_z= -\frac{u_x d_x u_x}{\omega_{ci} \cos \alpha} \left(1- \frac{c_s^2}{u_x^2}\right),
\end{equation}
and the electrostatic potential
\begin{equation}
\frac{e}{T_{e0}}(\phi(u_x)-\phi_0) = \ln \left(\frac{u_{x0}}{u_x}\right).
\end{equation}	

\bibliography{mag_presheath_grazing_incidence}

\end{document}